\documentclass[12pt]
{amsart}
\usepackage{amsmath}
\usepackage{amssymb,latexsym}
\usepackage{amscd}
\usepackage{verbatim}
\newcommand{\be}{\begin{equation}}
\newcommand{\bs}{\begin{sub}}
\newcommand{\es}{\end{sub}}
\newcommand{\bsn}{\begin{subn}}
\newcommand{\esn}{\end{subn}}
\newcommand{\bean}{\begin{eqnarray*}}
\newcommand{\eean}{\end{eqnarray*}}
\newcommand{\BA}[1]{\begin{array}{#1}}
\newcommand{\EA}{\end{array}}
\newcommand{\Real}{\mathbb{R}}

\newcommand{\NN}{\mathbb{N}}
\newcommand{\ZZ}{\mathbb{Z}}
\newcommand{\CC}{\mathbb{C}}
\newcommand{\RR}{\mathbb{R}}
\newcommand{\TT}{\mathbb{T}}

\newlength{\wex}  \newlength{\hex}
            \def\gl{\lambda}

\newcommand{\pf}{\noindent \mbox{{\bf Proof}: }}
\def\squarebox#1{\hbox to #1{\hfill\vbox to #1{\vfill}}}

\newcommand{\beq}{\begin{equation}}
\newcommand{\eeq}{\end{equation}}
\newcommand{\beqa}{\begin{eqnarray}}
\newcommand{\eeqa}{\end{eqnarray}}
\newcommand{\beqanl}{\begin{eqnarray*}}
\newcommand{\eeqanl}{\end{eqnarray*}}
\begin{document}
\renewcommand{\theequation}{\thesection.\arabic{equation}}
\newcommand{\mysection}[1]{\section{#1}\setcounter{equation}{0}}

\def\stackunder#1#2{\mathrel{\mathop{#2}\limits_{#1}}}
\newtheorem{theorem}{Theorem}
\newtheorem{lemma}[theorem]{Lemma}
\newtheorem{definition}[theorem]{Definition}
\newtheorem{corollary}[theorem]{Corollary}
\newtheorem{conjecture}[theorem]{Conjecture}
\newtheorem{remark}[theorem]{Remark}
\def\stackunder#1#2{\mathrel{\mathop{#2}\limits_{#1}}}

\author{Peter Kuchment}
\address{Department of Mathematics\\ Texas A\&M University\\
College Station, TX 77843-3368, USA}
\email{kuchment@math.tamu.edu}
\author{Yehuda Pinchover}
\address{Department of Mathematics\\ Technion - Israel Institute of
Technology\\ Haifa 32000, Israel}
\email{pincho@techunix.technion.ac.il}
\title[Liouville theorems on abelian coverings]%
{Liouville theorems and spectral edge behavior on abelian
coverings of compact manifolds}
\date{}
\begin{abstract}
The paper describes relations between Liouville type theorems for
solutions of a periodic elliptic equation (or a system) on an
abelian cover of a compact Riemannian manifold and the structure
of the dispersion relation for this equation at the edges of the
spectrum. Here one says that the Liouville theorem holds if the
space of solutions of any given polynomial growth is finite
dimensional. The necessary and sufficient condition for a
Liouville type theorem to hold is that the real Fermi surface of
the elliptic operator consists of finitely many points (modulo the
reciprocal lattice). Thus, such a theorem generically is expected
to hold at the edges of the spectrum. The precise description of
the spaces of polynomially growing solutions depends upon a
`homogenized' constant coefficient operator determined by the
analytic structure of the dispersion relation. In most cases,
simple explicit formulas are found for the dimensions of the
spaces of polynomially growing solutions in terms of the
dispersion curves. The role of the base of the covering (in
particular its dimension) is rather limited, while the deck group
is of the most importance.

The results are also established for overdetermined elliptic
systems, which in particular leads to Liouville theorems for
polynomially growing holomorphic functions on abelian coverings of
compact analytic manifolds.

Analogous theorems hold for abelian coverings of compact
combinatorial or quantum graphs.
\end{abstract}



\maketitle

\mysection{Introduction}\label{S:intro}

The classical Liouville theorem claims that any harmonic function
(i.e., a solution of the Laplace equation $\Delta u=0$) in $\RR^n$
that has a polynomial upper bound is in fact a (harmonic)
polynomial. In particular, the space of all harmonic functions
that grow not faster than $C(1+|x|)^N$, is of the finite
dimension\footnote{We will also use the notation
\begin{equation}
q_{n,N}:= \left(\begin{array}{c}n+N\\N\end{array}\right)
\label{qn}
\end{equation}
for the dimension of the space of all polynomials of degree at
most $N$ in $n$ variables. Notice that $q_{n-1,N}$ also coincides
with the dimension of the space of all homogeneous polynomials of
degree $N$ in $n$ variables, so in particular,
$h_{n,N}=q_{n,N}-q_{n,N-2}=q_{n-1,N-1}+q_{n-1,N}$.}
\begin{equation}\label{hn}
h_{n,N}:= \left(\begin{array}{c}n+N\\N\end{array}\right)-
\left(\begin{array}{c}n+N-2\\N-2\end{array}\right).
\end{equation}
The problem of extending this result to more general elliptic
operators and to Laplace--Beltrami operators on general Riemannian
manifolds of nonnegative Ricci curvature\footnote{Without a
condition on the curvature, the hyperbolic plane, where there is an
infinite dimensional space of bounded harmonic functions, provides a
counterexample.} has gained prominence since the work of S.~T.~Yau
\cite{Yau}. The questions asked concern the finite dimensionality of
the spaces of solutions of a prescribed polynomial growth, estimates
of (and in rare cases formulas for) their dimensions, and the
structure of these solutions. One can find recent advances, reviews,
and references in \cite{CM, L, LW}. In particular, Yau's conjecture
on the validity of the Liouville theorem for Riemannian manifolds of
nonnegative Ricci curvature was proven in full generality by
T.~H.~Colding and W.~P.~Minicozzi in \cite{CM} (see also references
to previous partial solutions in \cite{L, LW}).

In the flat situation, an amazing case was discovered by
M.~Avellaneda and F.-H.~Lin \cite{AL} and later was also studied
by J.~Moser and M.~Struwe \cite{MS}. In these papers the authors
dealt with polynomially growing solutions of second-order
divergence form elliptic equations
\begin{equation}\label{E:divergence}
Lu=-\sum\limits_{1 \leq i,j \leq n}(a^{i,j}(x)u_{x_i})_{x_j}=0
\end{equation}
with real coefficients that are periodic with respect to the
lattice $\ZZ^n$ in $\RR^n$. They obtained a comprehensive answer
for (\ref{E:divergence}) (see also \cite{CM,L} for related results
and references). Using the formalism of homogenization theory, it
was proved that the space of all solutions of the equation $Lu=0$
of polynomial growth of order at most $N$ has the same dimension
$h_{n,N}$ (see (\ref{hn})) as the space of harmonic polynomials in
$\RR^n$ of the same rate of growth.
Moreover, any solution $v$ of the equation $Lv=0$ in
$\mathbb{R}^n$ of polynomial growth is representable as a finite
sum of the form
\begin{equation}\label{polyn}
v(x)=\sum\limits_{j=(j_1,\ldots,j_n)\in \ZZ_{+}^n}x^jp_j(x),
\end{equation}
where the functions $p_j(x)$ are periodic with respect to the
group of periods of the equation and $x^j=\prod_ix_i^{j_i}$.

One can say that there has been no complete understanding of this
result concerning periodic equations. In particular, one can ask
the following natural questions:
\begin{itemize}
  \item[(i)]  Is it important that the operator is of divergence
  form?\\[-4mm]
  \item[(ii)] Can the results be generalized to higher order equations?
\\[-4mm]
  \item[(iii)] Is it possible to determine for a given periodic elliptic
equation whether the Liouville theorem holds?
\\[-4mm]
  \item[(iv)] How crucial is the usage of homogenization theory tools
(which automatically restricts the class of equations)?
\\[-4mm]
  \item[(v)] Same questions about elliptic systems.
\\[-4mm]
 \item[(vi)]Can these results be generalized for covering spaces of
compact manifolds?
\end{itemize}

Some partial answers to these questions were obtained in
\cite{KuPin, LW1}. In \cite{LW1}, the results for the divergence
type operators (\ref{E:divergence}) in $\RR^n$ were generalized to
the case of second-order periodic operators without lower order
terms. At the same time, \cite{KuPin} contains a necessary and
sufficient condition for the validity of the Liouville theorem for
a general periodic elliptic operator in $\RR^n$, as well as a
description (in most cases, implicit) of the dimensions of the
corresponding spaces of solutions. In particular, an explicit
formula was given for a general second-order periodic elliptic
operator that admits a global positive solution.

Simultaneously, Liouville type theorems for holomorphic functions on
complex analytic manifolds were studied (see
\cite{Brud,Brud2,Kaim,Lin,LinZaid1,LZ} and references therein). For
instance, one asks whether Liouville theorems for holomorphic
functions on coverings of compact analytic manifolds (or more
generally, on coverings of manifolds with the Liouville property)
hold true. One should mention the results of \cite{Lin}, where it
was shown in particular that nilpotent coverings of compact complex
analytic manifolds do have the Liouville property for {\em bounded}
holomorphic functions (i.e., the space of such functions is
finite-dimensional). It was not clear whether in these cases the
spaces of holomorphic functions of a given polynomial growth are of
finite dimensions as well. The exception was the result of
\cite{Brud}, where such a Liouville theorem was proven for abelian
coverings of compact K\"{a}hler manifolds (see also \cite{Brud2}).
This result also follows from \cite{CM}, while this relation with
Liouville theorems for harmonic functions disappears for
non-K\"{a}hler case.

One should also mention parallel studies concerning Liouville
theorems for harmonic functions on graphs (e.g.,
\cite{Kaim,Marg}).

The goal of this paper is to provide results of Liouville type that
clarify this issue for abelian covers of compact manifolds. The
results apply to elliptic equations and systems (including
overdetermined ones) on abelian coverings of compact Riemannian
manifolds, as well as to holomorphic functions on abelian coverings
of compact complex manifolds, and to periodic equations on abelian
coverings of combinatorial and quantum graphs. The crucial
techniques used in the paper are different from the ones used in all
the works cited above, with the exception of the authors' paper
\cite{KuPin}. The ideas and the techniques come from the Floquet
theory \cite{Ku,RS} and are related to some spectral notions common
in the solid state physics \cite{AM}. The reader can find all the
necessary preliminary information in the next two sections. In
comparison with \cite{KuPin}, the present paper provides explicit
formulas for the dimensions of the corresponding spaces, where
\cite{KuPin}, in general, contains only an implicit algorithm to
calculate these numbers. Moreover, multiplicities at spectral edges,
as well as overdetermined systems (in particular,
$\overline{\partial}$-operators and Liouville theorems for
holomorphic functions) are now allowed. Furthermore, operators on
graphs are also considered. The general theorems are applied to a
variety of specific examples of operators.

In order to outline the results of the paper, let us introduce
some objects. Consider a normal abelian covering\footnote{The word
`covering' in this paper always means `normal covering'.} of a
compact $d$-dimensional Riemannian manifold $M$
$$
X \mathop{\mapsto}\limits^{G} M,
$$
where $G$ is the (abelian) deck group of the covering. Without
loss of generality, one can assume that $G =\ZZ^n$. In fact, no
harm will be done if the reader imagines for simplicity that
$X=\RR^n$, $G =\ZZ^n$, and $M$ is the torus $\RR^n / \ZZ^n$
(albeit in general, the dimension $d$ of $M$ does not have to be
equal to $n$).

We will need to consider characters $\chi$ of $G$, i.e.,
homomorphisms of $G$ into the multiplicative group $\CC^*$ of
nonzero complex numbers. Unitary characters map $G$ into the group
$S^1$ of complex numbers of absolute value $1$. For any character
$\chi$, a function $f$ on $X$ will be called $\chi$-{\em
automorphic} if $f(gx)=\chi (g)f(x)$ for any $x\in X,\, g\in G $.

Let $P$ be an elliptic $G$-periodic operator on $X$ (in what
follows, we use for shortness the word `periodic' instead of
`$G$-periodic'). For any character $\chi$, consider the space of
$\chi$-automorphic functions on $X$. It can also be interpreted as
the space of sections of a linear bundle over $M$ determined by
$\chi$. It is invariant with respect to $P$, so one can consider
the restriction $P(\chi)$ of $P$ to this space\footnote{Exact
definitions of function spaces and operators are provided in the
next section.}. In the particular case of $\chi(g)\equiv 1$,
$P(1)$ is just the elliptic operator $P_M$ on $M$, whose lifting
to $X$ is $P$. In `non-pathological' cases, the spectra of all
operators $P(\chi)$ are discrete. The spectrum of $P(\chi)$, as a
multiple-valued function of the character $\chi$, is said to be
the {\em dispersion curve} or {\em dispersion relation}.

Let $N\geq 0$. We say that {\em the Liouville theorem of order $N$
for the equation $Pu=0$} holds true if the space $\mathrm{V}_N
(P)$ of solutions of the equation $Pu=0$ on $X$ that satisfy
$|u(x)|\leq C(1+\rho (x))^N$ for all $x\in X$ is of finite
dimension. Here $\rho (x)$ is the distance of $x\in X$ from a
fixed point $x_0 \in X$.

We can now formulate a general (and somewhat vague at this point)
statement that outlines our main results for the elliptic case
contained in theorems \ref{T:Liouville}, and
\ref{T:Liouville_dim}. The results for the overdetermined,
holomorphic, and graph cases can be found in Sections
\ref{S:overdet} and \ref{S:graphs}.

\vskip 3mm

\noindent{\bf Main Theorem.}{\em
\begin{enumerate}
\item If the Liouville theorem of an order $N\geq 0$ for the equation $Pu=0$ holds true,
then it holds for any order.

\item In order for the Liouville theorem to hold, it is necessary
and sufficient that the number of unitary characters $\chi$ for
which the equation $Pu=0$ has a nonzero $\chi$-automorphic
solution is finite.

\item If the Liouville theorem holds and the spectra of all the operators $P(\chi)$ are
discrete, then under some genericity conditions on the operator
$P$, the dimension of the space $\mathrm{V}_N (P)$ can be computed
in terms of the dispersion relation for $P$.

\item If the Liouville theorem holds, then all solutions that
belong to $\mathrm{V}_N (P)$ are linear combinations of Floquet
solutions of order $N$ (see the definitions in the text below).

\item Under the same conditions, one can describe a constant
coefficient (`homogenized') linear differential operator
$\Lambda(D)$ on $\RR^n$, such that there is a one-to-one
correspondence between the polynomial solutions of $\Lambda v=0$
on $\RR^n$ and the polynomially growing solutions of $Pu=0$ on
$X$.

\end{enumerate}
}

We will see that this result in particular means that one should
naturally expect the Liouville theorem to hold only when zero is
at an edge of the spectrum of $P$. This is true, for instance, for
the operators considered in \cite{AL,LW1,MS}, when zero is the
bottom of the spectrum.

It is interesting to notice that the dimension of $X$ does not have
to be equal to $n$, so the operators $P$ and $\Lambda$ might act on
manifolds of different dimensions. This happens since the Liouville
property is of a `homogenized' nature, i.e., it is something one
sees by looking at the manifold $X$ `from afar'. Thus, the local
details of the manifold are essentially lost and one sees the
euclidian space $\RR^n$ instead. In other words, the free rank of
the deck group of the covering plays a more prominent role for
Liouville theorems than the dimension of the manifold\footnote{This
resonates with M.~Gromov's notion of quasi-isometry, when the space
of a covering might be indistinguishable from its deck group
\cite{Hyperb,Gromov}.}.

Similar results hold for elliptic systems, including
overdetermined ones that are elliptic in the sense of being a part
of an elliptic complex of operators. The reader can find basic
notions and results concerning elliptic complexes in many books
and articles (e.g., \cite[Vol. III, Section 19.4]{Horm_differ},
\cite[Section 3.2.3]{Rempel}, \cite[Section IV.5]{Wells}). For the
particular case of the Cauchy--Riemann $\bar{\partial}$ operator,
one obtains a Liouville theorem for holomorphic functions on
abelian coverings (Theorem \ref{T:holom}). Analogs for operators
on combinatorial and quantum graphs are also straightforward to
obtain.

The outline of the paper is as follows. The next section
introduces necessary notations and preliminary results from the
Floquet theory, in particular the definition and properties of the
{\em Floquet--Gelfand transform}. The proofs are the same as for
the case of periodic operators on $\mathbb{R}^d$ and are hence
mostly omitted (see, e.g., how they can be worked out in parallel
to the flat case in \cite{KobSun} and also in
\cite{BrunSun,BrunSun2,Sunada}). The crucial Section
\ref{S:Floquet} is devoted to the detailed study of the so called
{\em Floquet--Bloch solutions}. In Section \ref{S:Liouv} we derive
Liouville type theorems for elliptic systems. Section
\ref{S:examples} provides some examples of applications to
specific periodic operators, and Section \ref{S:overdet} treats
overdetermined systems, including the case of analytic functions
on complex manifolds. Graphs are briefly considered in Section
\ref{S:graphs}.  The last sections contain concluding remarks and
acknowledgments.

\mysection{Notations and preliminary results}\label{S:notations}

Let us introduce first some standard notions of Floquet theory
(see \cite{Ea, Ku, RS}), which we will adjust to the case of
abelian covers (this does not require any change in the
substance).

Let $X$ be a noncompact smooth Riemannian manifold of dimension
$d$ equipped with an isometric, properly discontinuous, and free
action of a {\bf finitely generated abelian group} $G$. The action
of an element $g\in G$ on $x\in X$ will be denoted by $gx$.
Consider the orbit space $M=X/G$, which due to our conditions is a
Riemannian manifold of its own. {\bf We will assume that $M$ is
compact}. Hence, we are dealing with an abelian covering of a
compact manifold
\begin{equation}\label{E:cover0}
\pi:\,X \rightarrow M (=X/G).
\end{equation}
Switching to a subcovering $X \rightarrow \tilde{M}
\left(\rightarrow M\right)$ with a compact $\tilde{M}$, one can
eliminate the torsion part of $G$. In what follows, we could
substitute $\tilde{M}$ for $M$ and hence reduce the group $G$ to
$\mathbb{Z}^n$ with some $n\in \mathbb{N}$. {\bf We will therefore
assume from now on that} $G=\mathbb{Z}^n$. This will not reduce
the generality of the results.

Let $P$ be an elliptic operator of order $m$ on $X$ with smooth
coefficients\footnote{The smoothness condition can be
significantly reduced (see the corresponding remarks in Section
\ref{S:remarks}).} that commutes with the action of $G$. Such an
operator can be pushed down to an elliptic operator $P_M$ on $M$
(or conversely, $P$ is the lifting of $P_M$ to $X$). The
ellipticity is understood in the sense of the nonvanishing of the
principal symbol of the operator $P$ on the cotangent bundle (with
the zero section removed) $T^*X\setminus (X\times \{0\})$. The
dual operator (the formal adjoint) $P^{*}$ has similar properties
(in particular, $P^{*}$ is also $G$-periodic). Here the duality is
provided by the bilinear rather than the
sesquilinear\footnote{This is not essential, but simplifies
somewhat the calculations.} form
\[
<g,f>=\int\limits_{X}f(x)g(x)\,\mathrm{d}x.
\]

All the preparatory facts and main statements here hold for linear
periodic matrix operators that are either standard elliptic (some
times called elliptic in the Petrovsky sense) or elliptic in the
Douglis--Nirenberg sense (e.g., \cite{Douglis}, \cite[Vol. III,
Section 19.5]{Horm_differ}, and \cite[Section 3.1.2.1]{Rempel}).
The only difference in the proofs between the scalar and system
cases arises in the necessity of introducing appropriate spaces of
vector-valued functions, exactly as it was done in \cite[Section
3.4]{Ku}. Doing this, however, would on one hand be very routine,
and on the other hand would make reading the text more difficult.
Bearing this in mind, we will provide detailed considerations for
scalar linear elliptic operators only. They transfer with no
effort to systems.

For any {\em quasimomentum} $k \in \mathbb{C}^n$ we denote by
$\gamma_k$ the character of $G =\mathbb{Z}^n$ defined as
$\gamma_k(g)=\mathrm{e}^{\mathrm{i}k\cdot g}$. Here $g=(g_1,
\ldots ,g_n) \in \ZZ^n$ and $k\cdot g = k_1g_1+\cdots+k_ng_n$. We
will also use the notation $|g|=|g_1|+\cdots+|g_n|$ and for a
multi-index $j=(j_1,\ldots,j_n)\in \ZZ^n_+$ we denote $g^j=\prod_i
g_i^{j_i} $. If $k \in \RR^n$, the corresponding character is
unitary. Due to the obvious $2\pi$-periodicity of $\gamma_k$ with
respect to $k$, it is sufficient to restrict ourselves to the real
vectors $k$ in the {\it Brillouin zone} $\mathrm{B}=[-\pi ,\pi
]^n$, which is a fundamental domain of the reciprocal (dual)
lattice $G ^{*}=\left( 2\pi \ZZ\right) ^n$. Periodizing
$\mathrm{B}$ (i.e., considering the torus $\TT^n=\RR^n / 2\pi
\ZZ^n$), one obtains the dual group $\TT^n$ to $G =\ZZ^n$.

For any $k\in\CC^n$, we define the subspace $L^2_k(X)$ of
$L_{\mathrm{loc}}^2(X)$ consisting of all functions $f(x)$ that
are $\gamma_k$-automorphic, i.e., such that
$f(gx)=\gamma_k(g)f(x)=\mathrm{e}^{\mathrm{i}k\cdot g}f(x)$ for
a.e. $x\in X$. Alternatively, this space can be defined as
follows. We can identify $\gamma_k$ with a one-dimensional
representation of $G$ and consider the one-dimensional flat vector
bundle $E_k$ over $M$ associated with this representation. Then
elements of $L^2_k(X)$ can be naturally identified with
$L^2$-sections of $E_k$. A similar construction works also for
other classes of functions (e.g., from Sobolev spaces). We will
identify $G$-periodic (i.e., $\gamma_0$-automorphic) functions on
$X$ with functions on $M$. Due to the periodicity of the operator
$P$, it leaves the spaces $L^2_k$ invariant, and so its
restrictions to these subspaces define elliptic operators $P(k)$
on the spaces of sections of the bundles $E_k$ over $M$
\footnote{In the case when $X=\RR^n$ with the natural $\ZZ^n$
action, these operators can be identified with the ``shifted''
versions $P(x,D+k)$ of the operator $P$ acting on the torus
$\TT^n=\RR^n / \ZZ^n$ (see \cite{Ku,RS}).}. If $P$ is selfadjoint,
then $P(k)$ is selfadjoint for any real quasimomentum $k$.

It is natural that Fourier transform with respect to the
periodicity group $G$ reduces the space $L^2(X)$, as well as the
original operator $P$ on $X$, to the direct integral of operators
$P(k)$ on sections of $E_k$:
\begin{equation}\label{E:space_integral}
    L^2(X)=\int\limits^\bigoplus_\mathrm{B} L^2_k(X)\, \mathrm{d}k,
\end{equation}
and
\begin{equation}\label{E:operator_integral}
 P=\int\limits^\bigoplus_\mathrm{B} P(k)\,\mathrm{d}k.
\end{equation}
The integral is understood with respect to the normalized Haar
measure on the dual group $\TT^n$, which boils down to the
normalized Lebesgue measure $\mathrm{d}k$ on the Brillouin zone
$\mathrm{B}$. The isomorphism (in fact, an isometry) in
(\ref{E:space_integral}) is provided by an analog of the Fourier
transform (see \cite[Section 2.2]{Ku}, \cite{RS,Sunada}), which we
will call the {\it Floquet--Gelfand transform} ${\mathcal U}$:
\begin{equation}
f(x)\rightarrow {\mathcal U}f(k,x)=\sum_{g \in G
}f(gx)\gamma_{-k}(g) \qquad k \in \CC^n.  \label{Floquet}
\end{equation}
This transform is the main tool in the Floquet theory for PDEs
(e.g., \cite{Ku,RS,Skrig_book,Sunada}). It was introduced first in
\cite{Gelf} in order to obtain expansions into Bloch generalized
eigenfunctions for periodic selfadjoint elliptic operators.

It is not hard to describe the image of a Sobolev space $H^s(X)$
under the Floquet--Gelfand transform. In order to do so, let us
consider a quasimomentum $k\in \CC^n$ and denote by $H^s_{k}$ the
closed subspace of the space $H^s_{\mathrm{loc}}(X)$ consisting of
$\gamma_k$-automorphic functions. It is clear that $H^s_{k}$ can
be naturally equipped with the structure of a Hilbert space and
that it can be identified with the space $H^s(E_k)$ of
$H^s$-sections of the bundle $E_k$ over $M$.

One can show\footnote{See Theorem 2.2.1 in \cite{Ku} for the case
$X=\RR^n$. The general case of abelian covers over compact
manifolds is entirely parallel.} that
\begin{equation}
{\mathcal E}^s:=\mathrel{\mathop{\bigcup }\limits_{k\in \CC^n}}
H^s_{k}  \label{bundle}
\end{equation}
forms a holomorphic $2\pi \ZZ^n$-periodic Banach vector bundle. As
any infinite dimensional analytic Hilbert bundle over a Stein
domain, it is trivializable \cite{Bungart} (see also the survey
\cite{ZK} and theorems 1.3.2, 1.3.3, and 1.5.23 in \cite{Ku}).

We collect now several statements from \cite[Theorem XIII.97]{RS},
\cite[Theorem 2.2.2]{Ku}, and \cite{KuPin,Sunada}, recasted into the
abelian covering form:
\begin{theorem}\label{T:Plancherel}
\begin{enumerate}
\item  For any nonnegative integer $m$, the operator
$$
{\mathcal U}:H^{m}(X)\rightarrow L^2(\TT^n,{\mathcal E}^m)
$$
is an isometric isomorphism, where $L^2(\TT^n,{\mathcal E}^m)$ is
the space of square integrable sections over the torus (identified
with the Brillouin zone $\mathrm{B}$) of the bundle ${\mathcal
E}^m$, equipped with the natural topology of a Hilbert space.

\item Let $K\Subset X$ be a domain in $X$ such that
$\mathrel{\mathop{\cup }\limits_{g \in G}}gK=X$. Let also the space
\[
\mathrm{C}^m(X):=\!\left\{ \phi \!\in\! H_{\mathrm{loc}}^m(X)\mid
\;\sup_{g \in G }\left| \left| \phi \right| \right|
_{H^m(gK)}\!\!(1+\left| g \right|)^N\!\!<\!\infty \;\;\forall N
\right\}
\]
be equipped with the natural Fr\'{e}chet topology. Then
\[
{\mathcal U}:\mathrm{C}^m(X)\rightarrow C^\infty (\TT^n,{\mathcal
E}^m)
\]
is a topological isomorphism, where $C^\infty (\TT^n,{\mathcal
E}^m)$ is the space of $C^\infty $ sections of the bundle
${\mathcal E}^m$ over the complex torus $\TT^n$, equipped with the
standard topology.
%
%


\item Let the elliptic operator $P$ be of order $m$. Then under
the transform ${\mathcal U}$ the operator
$$
P: \mathrm{C}^{m}(X) \rightarrow \mathrm{C}^{0}(X)
$$
becomes the operator
\[
C^\infty (\TT^n,{\mathcal E}^m)\stackrel{P(k)}{\rightarrow }C^\infty (\TT^n,{\mathcal %
E}^0)
\]
of multiplication by the holomorphic Fredholm morphism $P(k)$
between the fiber bundles ${\mathcal E}^m$ and ${\mathcal E}^0$.

\end{enumerate}
\end{theorem}

\mysection{Floquet--Bloch solutions}\label{S:Floquet}

We now need to introduce and study our main notions: Bloch and
Floquet solutions of periodic differential equations.

\begin{definition}\label{D:Bloch}{\em
Let $k\in \CC^n$. A $\gamma_k$-automorphic function $u(x)$ on $X$
is said to be a {\em Bloch function with quasimomentum $k$}. In
other words, it is a function with the property
$u(gx)=\gamma_k(g)u(x)=\mathrm{e}^{\mathrm{i}k\cdot g}u(x)$ for
any $x\in X,\, g \in G$. To put it differently, $u(x)$ is
transformed according to an irreducible representation of the
group $G$ with the character $\gamma_k$.

A {\em Bloch solution} of an equation is a solution that is a
Bloch function.}
\end{definition}

Notice that every continuous Bloch function on $X$ with a real
quasimomentum (i.e., transformed according to an irreducible
unitary representation) is bounded. Any such Bloch function $u$
that belongs to $L^2_{\mathrm{loc}}(X)$ is bounded in the
following integral sense: for any compact $K\Subset X$ we have
$\sup_{g\in G}\|u\|_{L^2(gK)}<\infty$.

\vskip 2mm

 In the case when $X=\RR^n$, $G=\ZZ^n$, and $M=\TT^n$,
any Bloch function with quasimomentum $k$ has the form
$$
u(x)=\mathrm{e}^{\mathrm{i}k\cdot x}p(x)
$$
with a $\ZZ^n$-periodic function $p(x)$. In fact, a similar
(albeit less natural) representation holds for Bloch functions on
any abelian cover $X\mathop{\mapsto}\limits^{G}M$. Indeed, let $K$
be any fundamental domain of $X$ with respect to the action of $G$
and $f\in C_0^\infty(X)$ be a nonnegative function which is
strictly positive on $K$. We define for any $j=1,\ldots,n$
$$
h_j(x):=\sum_{g=(g_1,\ldots,g_n)\in G=\ZZ^n} f(gx)\exp(-g_j).
$$
Then $h_j(x)$ clearly is a positive function satisfying
$h_j(gx)=\mathrm{e}^{g_j}h_j(x)$ for any $g=(g_1,\ldots,g_n)\in
G$. It is an analog of $\mathrm{e}^{x_j}$ on $\RR^n$. Thus, one
can define analogs of powers $x^l=x_1^{l_1}\cdots x_n^{l_n}$ and
of exponents $\mathrm{e}^{\mathrm{i}k\cdot x}$ as follows: for
$l=(l_1,\ldots,l_n)\in \mathbb{Z}_+^n$ let
$$
[x]^l:= \prod_{j=1}^n \left[\log h_j(x)\right]^{l_j},
$$
and for any quasimomentum $k\in \mathbb{C}^n$
$$
e_k(x):=\exp(\mathrm{i}k_1\log h_1(x)+ \cdots +\mathrm{i}k_n\log
h_n(x)).
$$
Notice that $e_k(x)$ is a nonvanishing Bloch function on $X$ with
quasimomentum $k$, which is positive for $\mathrm{i}k\in
\mathbb{R}^n$. Thus, any Bloch function $u$ on $X$ with a
quasimomentum $k$ is given by
$$
u(x)=e_k(x)p(x),
$$
where $p(x)$ is $G$-periodic.

The construction of Bloch functions can be described in a more
invariant way \cite{Agmon_positive}. Consider a basis $\omega_j$ of
the space of all  closed differential $1$-forms on $M$ (modulo the
exact ones) such that their lifts $w_j$ to $X$ are exact. According
to De Rham's theorem, this basis is finite. One can now achieve the
same goals as before defining $h_j (x)=\exp (\int\limits_o^x w_j)$
for a fixed reference point $o\in X$.

We can now define a more general class than Bloch functions.

\begin{definition}\label{D:deffloq}{\em
A function $u(x)$ on $X$ is said to be a {\em Floquet function
with quasimomentum $k\in \CC^n$} if it can be represented in the
form
\begin{equation}\label{E:floquet_function}
u(x)=e_k (x)\left( \sum\limits_{\begin{array}{c}
j=(j_1,\ldots ,j_n)\in \ZZ%
_{+}^n\\
|j|\leq N
\end{array}}[x]^jp_j(x)\right),
\end{equation}
where the functions $p_j$ are $G$-periodic.

The number $N$ in this representation will be called {\em the
order} of the Floquet function.

A {\em Floquet solution} of an equation is a solution that is a
Floquet function.}
\end{definition}

This definition is modelled closely after the notion of Floquet
solution that is common in $\RR^n$ (e.g., \cite{Ku}), where the
formula is the same, one just replaces the `powers' $[x]^j$ by the
true powers $x^j$. However, unlike the notion of a Bloch function,
it lacks invariance. To alleviate this, we briefly address now a
different way to define Bloch and Floquet functions (solutions) on
abelian coverings \cite{Lin_priv}.

For any $g\in G$, let us define the first difference operator
$\Delta_g$ acting on functions on $X$ as follows:
\begin{equation}\label{E:difference_operator}
  \Delta_gu(x)=u(gx)-u(x).
\end{equation}
Clearly, $u$ is a periodic function (i.e., a Bloch function with
zero quasimomentum) if an only if it is annihilated by $\Delta_g$
for any $g\in G$. In fact, it is sufficient to check this property
for any set $\{g_j\}$ of generators of $G$. One wonders whether
one can check in a similar way whether a function is a Bloch
function with a nonzero quasimomentum and whether Floquet
functions allow for similar tests. In order to get the answer, we
need to introduce a twisted version of the first difference, that
depends of the quasimomentum:
\begin{equation}\label{E:twisteddifference_operator}
  \Delta_{g;k}u(x)=\chi_{-k}(g)u(gx)-u(x)=\mathrm{e}^{-\mathrm{i}k\cdot g}u(gx)-u(x).
\end{equation}

We also need to introduce iterated finite differences of order $N$
with quasimomentum $k$ as follows:
\begin{equation}\label{E:iterated_differences}
\Delta_{g_1,\ldots,g_N;k}=\Delta_{g_1;k}\cdot \ldots \cdot
\Delta_{g_{N};k},
\end{equation}
where $g_j\in G$ (it will always be sufficient to use only
elements (maybe repeated) of a fixed set of generators of $G$).

We can now answer the question by proving the following
\begin{lemma}
A function $u(x)$ on $X$ is a Floquet function of order $N$ with
quasimomentum $k$ if and only if it is annihilated by any
difference of order $N+1$ with quasimomentum $k$:
$$
\Delta_{g_1,\ldots,g_{N+1};k}u=0\qquad \forall g_1,\ldots,g_{N+1}
\in G,
$$
(choosing $g_j$ from a fixed set of generators is sufficient). In
particular, it is a Floquet function of order $N$ with
quasimomentum $0$ if and only if
\begin{equation}\label{E:difference_condition}
\Delta_{g_1}\cdot \ldots \cdot \Delta_{g_{N+1}}u=0 \qquad \forall
g_1,\ldots,g_{N+1} \in G.
\end{equation}
\end{lemma}
\pf Let us provide the proof for the case $k=0$ first. As it has
already been mentioned, the necessity and sufficiency of the
condition (\ref{E:difference_condition}) checks out easily for
$N=0$, where it boils down to $\Delta_gu=0$ for all $g \in G$,
i.e., to the periodicity of $u$. Necessity for any $N$ now follows
easily by induction if one takes into account the representation
(\ref{E:floquet_function}). Indeed, one computes that (using the
same standard basis $\{g_j\}$ of $\ZZ^n$ as before)
\begin{equation}\label{E:diff_powers}
\Delta_{g_j}[x]^{(l_1,\ldots,l_n)}=l_j[x]^{(l_1,\ldots,l_j-1,\ldots,l_n)}+\mbox{
lower order terms}.
\end{equation}
Here `lower order terms' contain linear combinations of $[x]^l$s
of strictly lower total degrees. Since the difference operators do
not alter periodic functions, we obtain that any $\Delta_{g_j}$
reduces Floquet functions of order $N+1$ to the ones of order $N$,
which concludes the induction step of the proof of necessity for
$k=0$.

Let us prove sufficiency which also follows by induction with
respect to the order of the Floquet function. We have already
checked it for $N=0$. Assume that this has been proven for orders
up to $N$. Suppose that $u$ satisfies $\Delta_{g_1}\cdot \ldots
\cdot \Delta_{g_{N+2}}u=0$ for any $g_1,\ldots,g_{N+2} \in G$.
Take a set of generators $\{g_j\}_{j=1}^n$ in $G$. Then the
functions $f_j:=\Delta_{g_j}u$ satisfy the condition of the Lemma
for the order $N$. According to the induction hypothesis, we
conclude that
$$
f_j(x)=\left( \sum\limits_{\begin{array}{c}
l=(l_1,\ldots ,l_n)\in \ZZ%
_{+}^n\\
|l|\leq N
\end{array}}[x]^lp_{l,j}(x)\right)
$$
with periodic functions $p_{l,j}$.

We claim that for any Floquet function $f$ of order $N$ (with
$k=0$) there exists a Floquet function $v$ of order $N+1$ such
that $\Delta_{g_j}v=f$. Without loss of generality we assume that
$j=1$. Since the difference operator $\Delta_{g_1}$ does not
change periodic coefficients, it is sufficient to check the
statement for $f=[x]^l$, where $|l|\leq N$. Now induction with
respect to $l_{1}$ finishes the job. Namely, (\ref{E:diff_powers})
provides a linear system of a triangular structure for recursively
determining the coefficients of $v$ such that $\Delta_{g_j}v=f$.

Therefore,  a Floquet function $v_1$ of order $N+1$ exists such
that $\Delta_{g_1}v_1=\Delta_{g_1}u$. This means that the function
$u_2=u-v_1$ is periodic with respect to the one-parameter subgroup
generated by $g_1$. Since the condition of the Lemma is still
satisfied for $u_2$, and since $u_2$ is $g_1$-periodic, we can
continue this process and find a Floquet function $v_2$ of order
$N+1$ that is $g_1$-periodic and such that $u_3=u-v_1-v_2$ is
periodic with respect to both $g_1$ and $g_2$. Continuing this
process, we get the conclusion of the Lemma for $k=0$.

Let us now prove the statement for an arbitrary quasimomentum $k$.
According to the definition (\ref{E:floquet_function}) of Floquet
functions, any Floquet function $u$ of order $N$ with
quasimomentum $k$ has the form $u=e_k(x)v$, where $v$ is a Floquet
function of order $N$ with quasimomentum $k=0$. A straightforward
calculation shows that if any two functions $u$ and $v$ satisfy
$u=e_k(x)v$, then one has $\Delta_{g;k}u=e_k(x)\Delta_{g}v$. Thus,
the statement of the lemma for an arbitrary quasimomentum $k$
follows from the one for $k=0$. \qed

We have provided several different ways to interpret the notion of
Floquet solutions: using explicit formulas analogous to the ones
in $\RR^n$ (constructing analogs of coordinate functions and
exponents by either explicit constructions, or by using some
special differential forms on $M$), as well as in terms of some
difference operators\footnote{Difference operators approach has
been successfully used in related studies of periodic equations
and Liouville type problems for holomorphic functions in
\cite{Brud,Lin,LP,MS}.}. Another way to think of Floquet functions
is to imagine the finite dimensional subspace $E$ generated by the
$G$-shifts of such a function as a Jordan block for the action of
$G$ \cite{Pa2}. This suggests a relation to indecomposable
(non-unitary) representations of $G$ (e.g., \cite[Ch. 6, Sect.
3]{Barut}). However, we do not pursue this approach, due to the
known difficulties of classifying such representations
\cite{Gelf-Ponom}.

\begin{remark}\label{R:Floquet_grows}{\em
Any continuous Floquet function $u(x)$ of order $N$ with a real
quasimomentum satisfies the growth estimates $$|u(x)|\leq
C(1+|g|)^N \qquad \forall g\in G \mbox{ and } x\in gK,$$ where $C$
depends on $K$ and $u$. In general, one needs to replace this
growth estimate by an integral one, as it was done before for
Bloch functions.}
\end{remark}

We now introduce a notion that plays a very important role in
solid state physics, photonic crystal theory, as well as in the
general theory of periodic PDEs
\cite{AM,Ku,Ku-chapter,Novikov_VINITI}. It will be also crucial
for formulation of our main results.

\begin{definition} \label{defFermi}{\em
The (complex) {\em Fermi surface} $F_P$ of the operator $P$ (at
the zero energy level) consists of all quasimomenta $k\in \CC^n$
such that the equation $Pu=0$ on $X$ has a nonzero Bloch solution
with a quasimomentum $k$ .
 The {\em real Fermi surface} $F_{P,\RR}$ is $F_P \cap \RR^n$.}
\end{definition}
Equivalently, $k\in F_P$ means the existence of a nonzero solution
$u$ of the equation $P(k)u=0$. In the Euclidean case, such a $u$
has a form  $u(x)=\mathrm{e}^{\mathrm{i}k\cdot x}p(x)$, where
$p(x)$ is a $G $-periodic function. Fermi surface plays in the
periodic situation the role of the characteristic variety (the set
of zeros of the symbol) of a constant coefficient differential
operator.

Introducing a spectral parameter $\lambda$, one arrives at the
notion of the {\it Bloch variety}:
\begin{definition} \label{defBloch}{\em
The (complex) {\em Bloch variety} $B_P$ of the operator $P$
consists of all pairs $(k,\lambda) \in \CC^{\, n+1}$ such that the
equation $Pu=\lambda \, u$ has a nonzero Bloch solution $u$ with a
quasimomentum $k$. The {\em real Bloch variety} $B_{P,\RR}$ is
$B_P\cap \RR^{n+1}.$}
\end{definition}

The Bloch variety $B_P$ can be treated as the graph of a
(multivalued) function $\lambda(k)$, which is called the {\it
dispersion relation}. If the spectra of the operators $P(k)$ on
$M$ are discrete, we can single out continuous branches
$\lambda_j$ of this multivalued dispersion relation. They are
called the {\it band functions} \cite{RS, Ku}. The Fermi surface
is obviously the zero level set of the dispersion relation.

In order to justify the notion of a band function, we need to
guarantee the discreteness of the spectrum of the operators $P(k)$
on $M$ for all $k\in \mathbb{C}^n$. In other words, we need to
exclude the pathological (but possible) situation of the spectrum
of $P$ covering the complex plane. One of the reasons for such
strange spectral behavior is the Fredholm index being not equal to
zero. However, even when the index is equal to zero, such
pathology can occur, as the example of the operator
$\mathrm{e}^{\mathrm{i}\varphi}\mathrm{d}/\mathrm{d}\varphi$ on
the unit circle shows. Self-adjointness of $P$ is one of the
conditions that would obviously guarantee discreteness. Another
example is a second-order elliptic periodic operator in $\RR^n$ of
the form
\begin{equation}
L=-\sum_{i,j=1}^n a_{ij}(x)\partial _{i}\partial _{j}+\sum_{i=1}^n
b_i(x)\partial _i+c(x) \label{E:operator}
\end{equation}
with real and smooth coefficients. More sufficient conditions can
be found for example in \cite{Ag}.

\begin{lemma}\label{L:Fermi}
The Fermi and Bloch varieties are the sets of all zeros of entire
functions of a finite order in $\CC^{\, n}$ and $\CC^{\, n+1}$,
respectively.
\end{lemma}
This is proven in \cite[theorems 3.1.7 and 4.4.2]{Ku} for the flat
case. The case of a general abelian covering does not require any
change in the proof.

The lemma implies in particular, that the band functions
$\lambda(k)$ are piecewise-analytic (e.g., when there is no level
crossing, one has analyticity due to the standard perturbation
theory \cite{Kato}). This statement was originally proven in
\cite{Wilcox} for Schr\"{o}dinger operators.

Another useful property of the Bloch and Floquet varieties is the
relation between the corresponding varieties of the operators $P$
and $P^*$:
\begin{lemma}
\label{dual} \cite[Theorem 3.1.5]{Ku} A quasimomentum $k$ belongs
to $F_{P^*}$ if and only if $-k \in F_{P}$. Analogously,
$(k,\lambda)\in B_{P^*}$ if and only if $(-k,\lambda)\in B_{P}$.
In other words, the dispersion relations $\lambda(k)$ and
$\lambda^*(k)$ for the operators $P$ and $P^*$ are related as
follows:
\begin{equation}
\lambda^*(k)=\lambda(-k).
\label{dual_disp}
\end{equation}
\end{lemma}




We will need to see how the structure of the functions of Floquet
type (see Definition \ref{D:deffloq}) and in particular, of
Floquet solutions of our periodic equation reacts to the
Floquet--Gelfand transform. For instance, in the constant
coefficient case, where the role of the Floquet solutions is
played by the exponential polynomials
$$
u(x) = \mathrm{e}^{\mathrm{i}k\cdot x}\sum\limits_{\left| j\right|
\leq N}p_jx^j,
$$
where $p_j\in \mathbb{C}$, such functions are Fourier transformed
into distributions supported at the point $\left(-k\right)$. The
next statement shows that under the Floquet--Gelfand transform,
each Floquet type function (\ref{E:floquet_function}) corresponds,
in a similar way, to a (vector valued) distribution supported at
the quasimomentum $\left( -k\right) $. This, and some other
properties of Floquet solutions that play a crucial role in
establishing our Liouville type theorems are collected in the next
lemma.

Every Floquet type function $u$ with a real quasimomentum is of
polynomial growth, and thus determines a (continuous linear)
functional on the previously defined space
$\mathrm{C}^0(X)$ (see Theorem \ref{T:Plancherel}). If it
satisfies the equation $Pu=0$ for a periodic elliptic operator of
order $m$, then as such a functional it is orthogonal to the range
of the dual operator
$P^{*}:\mathrm{C}^m(X)\rightarrow \mathrm{C}^0(X)$. According to
Theorem \ref{T:Plancherel}, after the Floquet--Gelfand transform
any such functional becomes a functional on $C^\infty\left(
\TT^n,{\mathcal E}^0\right)$ that is orthogonal to the range of
the operator of multiplication by the Fredholm morphism
$P^{*}(k):{\mathcal E}^m\rightarrow {\mathcal E}^0$. The following
auxiliary result (see \cite{KuPin}) describes all such
functionals\footnote{Although the discreteness of the spectrum of
$P(k)$ was assumed throughout the whole paper \cite{KuPin}, it is
in fact not needed for (and was not used in the proof of) this
lemma.}.
\begin{lemma}\label{L:Floq_struct}
\begin{enumerate}
\item A continuous linear functional $u$ on $\mathrm{C}^0(X)$ is generated by
a Floquet type function with a quasimomentum $k_0$ if and only if
after the Floquet--Gelfand transform it corresponds to a
functional on $C^\infty \left( \TT^n,{\mathcal E}^0\right)$ which
is a distribution $\phi $ that is supported at the point $-k_0$,
i.e., has the form
\begin{equation}\label{E:functional}
\left\langle \phi ,f\right\rangle =\sum_{\left| j\right| \leq
N}\left( q_j,\left. \frac{\partial ^{\left| j\right| }f}{\partial
k^j}\right| _{-k_0}\right)_{L^2(M)}  \quad f\in C^\infty \left(
\TT^n,{\mathcal E}^0\right),
\end{equation}
where $q_j\in L^2(M)$. The orders $N$ of the Floquet function and
of the corresponding distribution $\phi $ are the same.

\item Let $a_k$ be the dimension of the kernel of the operator
$$
P(k):H^m_k\rightarrow L^2_k.
$$
Then the dimension of the space of Floquet solutions of the
equation $Pu=0$ of order at most $N$ with a quasimomentum $k$ is
finite and does not exceed $a_{k}q _{n,N}$.
\end{enumerate}
\end{lemma}
The estimate on the dimension given in the Lemma
\ref{L:Floq_struct} is very crude and can often be improved. The
next result (Theorem \ref{T:Fl_dimen}) provides in some cases
exact values of these dimensions. This theorem is the crucial part
of the proof of (Liouville) Theorem \ref{T:Liouville}. In order to
state and prove it, we need to introduce some notions.

First of all, the analytic Hilbert bundle $\mathcal{E}^m$ is
locally trivial for any $m$. Since the previous Lemma shows that
our interest in Floquet solutions is local with respect to the
quasimomentum $k$, we can trivialize the bundles and hence assume
that the analytic families of operators $P(k)$ and $P^*(k)$ act in
a fixed Hilbert space. At this moment, we will need the additional
condition that the spectra of these operators are discrete (see
the corresponding discussion earlier in the text). Assume now that
zero is an eigenvalue of the adjoint operator
$P^*(-k_{0}):H^m_{-k_0}(X)\rightarrow L^2_{-k_0}(X)$ (since under
the conditions we imposed on the operator, its Fredholm index is
zero, this means that the operator $P(k_{0})$ has eigenvalue zero
as well). Suppose further that the algebraic multiplicity of the
null eigenvalue is equal to $r$. Consider a closed curve
$\Upsilon$ in $\mathbb{C}$ separating $0$ from the rest of the
spectrum of $P^*(-k_{0})$ and the corresponding (analytically
depending on $k$ in a neighborhood of $k_0$) $r$-dimensional
spectral projector $\Pi (k)$ for $P^*(-k)$. Let  $\{e_j\}_{j=1}^r$
be an orthonormal basis of the spectral subspace of $P^*(-k_{0})$
corresponding to the point $0\in\CC$ (i.e., the range of  $\Pi
(k_0)$). Some of our main results will be expressed in terms of
the following $r\times r$ matrix function
\begin{equation}\label{E:lambda-def}
\lambda(k)_{ij}=\langle e_j,(P^*(k)\Pi (k)e_i\rangle.
\end{equation}
This matrix function is analytic with respect to $k$ in a
neighborhood of $k_0$.
\begin{remark}\label{matrix_functions}
{\em We would like to mention that in what follows, the results will
be invariant with respect to a multiplication of the matrix
$\lambda$ from either side by an invertible matrix-function analytic
in a neighborhood of $k_0$. This means, in particular, that if
$r=1$, then $\lambda(k)$ can equivalently be chosen to be equal to
the analytic branch around $k_0$ of the eigenvalue of $P^*(-k)$ that
vanishes at $k_0$.

Moreover, instead of $\{\Pi (k)e_i\}$ in (\ref{E:lambda-def}) one
can use any holomorphic basis $e_j(k)$ in the range of $\Pi (k)$
near $k=k_0$. One can also use instead of $\langle
e_j,\bullet\rangle$ any family $\{f_j(k)\}_{j=1}^r$ of holomorphic
with respect to $k$ (in a neighborhood of $k_0$) functionals that is
complete on the range of $\Pi(k)$.}
\end{remark}

Consider the Taylor expansion of $\lambda(k)$ around the point
$k_0$ into homogeneous matrix-valued polynomials:
\begin{equation}\label{E:Taylor}
\lambda(k)=\sum\limits_{l \geq 0}\lambda_l(k-k_0).
\end{equation}
In this paper we will be mostly concerned with the first nonzero
term $\lambda_{l_0}$ of the expansion.

\begin{definition}\label{Qharm} {\em Let $Q$ be a homogeneous polynomial
in $n$ variables with matrix coefficients of dimension $r\times
r$, and let $Q(D)$ be the differential matrix operator with the
symbol $Q$. A $\mathbb{C}^r$-valued polynomial $p(x)$ in $\Real^n$
is called {\em $Q$-harmonic} if it satisfies the system of
differential equations $Q(D)p=0$.}
\end{definition}

Let ${\mathcal P}$ denote the vector space of all
$\mathbb{C}^r$-valued polynomials in $n$ variables, and let
$\mathrm{P}_l$ be the subspace of all such homogeneous polynomials
of degree $l$. So,
\begin{equation}\label{E:P}
{\mathcal P}=\bigoplus\limits_{l=0}^\infty \mathrm{P}_l
\end{equation}
and
\begin{equation}\label{E:PN}
{\mathcal P}_N:=\bigoplus\limits_{l=0}^N\mathrm{P}_l
\end{equation}
is the subspace of all such vector valued polynomials of degree at
most $N$. If $Q(k)$ is a homogeneous polynomial of degree $s$ with
values in $r \times r$ matrices, then the matrix differential
operator $Q(D)$ maps $\mathrm{P}_{l+s}$ to $\mathrm{P}_l$. If the
determinant $\det Q$ is not identically equal to zero, then this
mapping is surjective for any $l$ (this will follow from the proof
of the theorem below). Hence, the mapping
$Q(D):\mathcal{P}\rightarrow \mathcal{P}$ has a (non-uniquely
defined) linear right inverse $R$ that preserves the homogeneity
of polynomials.

\begin{theorem}
\label{T:Fl_dimen} Assume that zero is an isolated eigenvalue of
algebraic multiplicity $r$ of the operator $P^*(-k_{0}):
H^m_{-k_{0}}(X)\rightarrow L^2_{-k_{0}}(X)$,.  Let $\lambda (k)$
be defined in a neighborhood of $k_0$ as in
(\ref{E:lambda-def})\footnote{Any analytic function in a
neighborhood of $k_0$ that differs from $\lambda(k)$ by a left and
right multiplication by analytic invertible matrix functions will
produce the same results in what follows.}. Let also $\lambda
_{l_0}$ be the first nonzero term of the Taylor expansion
(\ref{E:Taylor}). Then
\begin{enumerate}
\item For any $N\geq 0$, the dimension of the space of Floquet
solutions of the equation $Pu=0$ in $X$ with quasimomentum $k_0$
and of order at most $N$ is finite and does not exceed $rq_{n,N}$.

\item If $\det \lambda_{l_0}$ is not identically equal to zero
(for instance, this is the case when the eigenvalue is simple,
i.e., $r=1$), then for any $N\geq 0$ the dimension of the space of
Floquet solutions of the equation $Pu=0$ in $X$ of order at most
$N$ and with quasimomentum $k_0$ is equal
to\begin{equation}\label{E:dim_Floquet}
  r\left[  \left(%
\begin{array}{c}
  n+N \\
  N\\
\end{array}%
\right) - \left(%
\begin{array}{c}
  n+N-l_0 \\
  N-l_0\\
\end{array}%
\right) \right].
\end{equation}

\item This dimension coincides with the dimension of the space of
all $\lambda _{l_0}$-harmonic polynomials of degree of at most $N$
with values in $\mathbb{C}^r$. Moreover, given a linear right
inverse $R$ of the mapping $\lambda
_{l_0}(D):\mathcal{P}\rightarrow \mathcal{P}$ that preserves
homogeneity, one can construct an explicit isomorphism between the
corresponding spaces.
\end{enumerate}
\end{theorem}
\pf As before, we will assume that the bundles are analytically
trivialized around the point $k_0$ and hence all operators act
between fixed spaces that we will denote $H^m$ and $H^0$. In order
to simplify notations, we assume that $k_0=0$ (this can always be
achieved by a change of variables). Let us denote by $N(k)$ the
range of the projector $\Pi (k)$ and choose a closed complementary
subspace $M$ to $N(0)$ in $ H^m$. The subspace $M$ stays
complementary to $N(k)$ in a neighborhood of $0$ and so
$$ H^m=M\oplus N(k). $$
 Thus, $P^{*}(-k)$ has zero kernel on $M$ for all $k$
in a neighborhood of $0$. This implies that the range of $P^*(-k)$ on $M$
forms an analytic Banach vector bundle $R(k)$ in a neighborhood of
$0$ (e.g., Theorem 1.6.13 of \cite{Ku}). Representing now the
operator $P^{*}(-k)$ in the block form according to the
decompositions
$$ H^m=M\oplus N(k)$$
and
$$
H^0=R(k)\oplus N(k), $$
we get
$$ P^{*}(-k)= \left(
\begin{array}{cc}
B(k)&0 \\
0&\tilde{\lambda} (-k)
\end{array}
\right), $$ where $B(k)$ is an analytic invertible
operator-function and the matrix analytic function
$\tilde{\lambda} (-k)$ differs from $\lambda (-k)$ only by
multiplying by an invertible analytic matrix function and thus for
our purposes can be replaced by the latter one.

Let us now have a functional $\phi$ on $C^\infty \left(
\TT^n,{\mathcal E}_0\right)$ supported at $0$, such that it is
orthogonal to the range of the operator of multiplication by
$P^{*}(-k)$. Then it must be equal to zero on all sections of the
bundle $R(k)$ (since they are all in the range, due to
invertibility of $B(k)$). This means that the restriction of such
functionals to the sections of the finite-dimensional bundle
$N(k)$ is an one-to-one mapping. This reduces the problem to the
following: find the dimension of the space of all distributions of
order $N$ supported at the origin such that they are orthogonal to
the sub-module generated by the matrix $\lambda (-k)$ in the
module of germs of analytic vector valued functions. One can
change variables to eliminate the minus sign in front of $k$. Due
to the finiteness of the order of the distribution, the problem
further reduces to the following: find the dimension of the
cokernel of the mapping
$$
\Lambda_N
:{\mathcal P}_N\rightarrow {\mathcal P}_N\,.
$$
Here $\Lambda_N(p)$ for $p\in {\mathcal P}_N$ is the Taylor
matrix-valued polynomial of order $N$ at $0$ of the product
$\lambda (k)p(k)$. Let us write the block matrix $\Lambda_{ij}$ of
the operator $\Lambda_N $ that corresponds to the decomposition
${\mathcal P}_N=\bigoplus\limits_{l=0}^N\mathrm{P}_l$. Then
$\Lambda _{ij}=0$ for $i-j<l_0$. For $i-j\geq l_0$ the entry
$\Lambda _{ij}$ is the operator of multiplication by $\lambda
_{i-j}$ acting from $\mathrm{P}_j$ into $\mathrm{P}_i$. Since
$\mbox{det}(\lambda_{l_0})$ is not identically equal to zero, it
follows that for $i-j= l_0$ the operator $\Lambda _{ij}$ of
multiplication by $\lambda_{l_0}$ has zero kernel.

In order to prove the theorem, we need to find the dimension of
the cokernel of $\Lambda_N$, as well as to obtain the cokernel's
description.

The first statement of the theorem is now obvious, since the
dimension of the cokernel of $\Lambda_N$ cannot exceed the
dimension of the ambient space $\mathcal{P}_N$, which is equal to
$rq_{n,N}$.

Let us now approach the second statement. Since $\Lambda_N$ is a
square matrix, $\mbox{ dim} \mbox{ Coker} \Lambda_N=\mbox{ dim}
\mbox{ Ker} \Lambda_N$. The latter dimension, however, is easy to
find, due to the triangular structure of the equation $\Lambda_N
p=0$, if it is written in the block matrix form according to the
decomposition (\ref{E:PN}): \begin{equation}\label{E:matrix}
    \left(%
\begin{array}{ccccccc}
  0 & \ldots &0 & \lambda_{l_0} & \lambda_{l_0+1} & \ldots & \lambda_{N}\\
  0 & \ldots &0 & 0& \lambda_{l_0} & \ldots & \lambda_{N-1}\\
  \vdots &\vdots & \vdots& \vdots &  & \vdots & \vdots\\
  0& \ldots &  & & & 0 & \lambda_{l_0} \\
 0 & \ldots &  & &   & & 0 \\
 \vdots& \vdots & \vdots & \vdots & \vdots & \vdots &\vdots\\
 0 & \ldots & &  &  &   & 0 \\
\end{array}%
\right)
\left(%
\begin{array}{c}
  p_N \\
  p_{N-1} \\
  \vdots \\
  p_{N-l_0} \\
  p_{N-l_0-1} \\
  \vdots \\
    p_0\\
\end{array}%
\right) =0.
\end{equation}
\noindent Here $p=p_0+\cdots+p_N$ is the expansion of $p\in
{\mathcal P}_N$ into homogeneous terms. Since $\mbox{
det}\lambda_{l_0}\neq 0$, one concludes immediately from
(\ref{E:matrix}) that $p_0=\ldots=p_{N-l_0}=0$, while other
components are arbitrary. This gives the dimension of the kernel
(and hence of the cokernel) of $\Lambda_N$ as
$$
r\left[  \left(%
\begin{array}{c}
  n+N \\
  N\\
\end{array}%
\right) - \left(%
\begin{array}{c}
  n+N-l_0 \\
  N-l_0\\
\end{array}%
\right) \right],
$$
which proves the second statement of the theorem.

In order to prove the last assertion of the theorem, we need
describe the elements of the cokernel of $\Lambda_N$. Hence, we
need to find the kernel of the adjoint matrix $\Lambda_N ^{*}$.
The adjoint matrix acts in the space
$\bigoplus\limits_{l=0}^N\mathrm{P}_l^{*}$, where
$\mathrm{P}_l^{*}$ can be naturally identified with the space of
linear combinations with coefficients in $\CC^r$ of the
derivatives of order $l$ of the Dirac's delta-function at the
origin. Here we have $\Lambda _{ij}^{*}=0$ for $j-i<l_0$, and for
$j-i\geq l_0$ the entry $\Lambda _{ij}^{*}$ is the dual to the
operator of multiplication by $\lambda _{j-i}$ acting from
$\mathrm{P}_i$ into $\mathrm{P}_j$. In particular, since for $j-i=
l_0$ the latter operator is injective, we conclude that the
operators $\Lambda _{ij}^{*}$ are surjective. This enables one to
describe the structure of the kernel of $\Lambda_N ^{*}$ (and
hence of cokernel of $\Lambda_N$). Namely, let
$$ \psi =\left( \psi _0,\ldots,\psi _N\right) \in
\bigoplus\limits_{l=0}^N\mathrm{P}_l^{*} $$ be such that $\Lambda
^{*}_N\psi =0$. Due to the triangular structure of $\Lambda_N
^{*}$, we can solve this system:
$$ \sum\limits_{j\geq i+l_0}\Lambda _{ij}^{*}\psi
_j=0\qquad i=0,\ldots,N-l_0. $$ Taking the Fourier transform, we
can rewrite this system in the form $$ \sum\limits_{j\geq
i+l_0}\lambda _{j-i}(D)\widehat{\psi _j}=0\qquad i=0,\ldots,N-l_0,
$$ where $\widehat{\psi}$ denotes the Fourier transform of $\psi$.
Therefore, $\widehat{\psi _j}$ is a homogeneous polynomial of
degree $j$ in $\Real^n$. For $i=N-l_0$ we have $$\lambda
_{l_0}(D)\widehat{\psi _N}=0. $$ This equality means that
$\widehat{\psi _N}$ can be chosen as an arbitrary $\lambda
_{l_0}$-harmonic homogeneous polynomial of order $N$ (with values
in $\mathbb{C}^r$). Moving to the previous equation, we
analogously obtain
$$ \lambda _{l_0}(D)\widehat{\psi _{N-1}}+\lambda
_{l_0+1}(D)\widehat{\psi _N}=0, $$ or $$
 \lambda _{l_0}(D)\widehat{\psi _{N-1}}=-\lambda _{l_0+1}(D)\widehat{\psi _N}.
$$ The right hand side is already determined, and the
nonhomogeneous equation, as we concluded before, always has a
solution, for instance $$
 -R\left( \lambda _{l_0+1}(D)\widehat{\psi _N}\right) .
$$
This means that
$$
\widehat{\psi _{N-1}}+R\left( \lambda _{l_0+1}(D)\widehat{\psi
_N}\right)
$$
 is a $\lambda _{l_0}$-harmonic homogeneous polynomial of order $N-1$. We see that
the solution $\widehat{\psi _{N-1}}$ exists and is determined up
to an addition of any homogeneous $\lambda _{l_0}$-harmonic
polynomial of degree $N-1$. Continuing this process until we reach
$\widehat{\psi _{0}}$, we conclude that the mapping $$ \psi
=\left( \psi _0,\ldots,\psi _N\right) \rightarrow \phi =\left(
\phi _0,\ldots,\phi _N\right), $$ where $$ \phi _j=\widehat{\psi
_j}+R\sum\limits_{i>j}\lambda _{i-j+l_0}(D)\widehat{\psi _i} $$
establishes an isomorphism between the cokernel of the mapping
$\Lambda_N $ and the space of $\lambda _{l_0}$-harmonic
polynomials of degree at most $N$. This proves the theorem. \qed

In the simplest possible cases, the theorem immediately implies
the following:
\begin{corollary}
\label{specific} Under the hypotheses of Theorem \ref{T:Fl_dimen}
one has:
\begin{enumerate}
\item If $k_0$ is a noncritical point of a single band function
$\lambda (k)$ such that $\lambda(0)=0$, then the dimension of the
space of Floquet solutions of the equation $Pu=0$ on $X$ of order
at most $N$ with a quasimomentum $k_0$ is equal to $q_{n-1,N}$,
i.e., it is equal to the dimension of the space of all polynomials
of degree at most $N$ in $\Real^{n-1}$.

\item If the Taylor expansion at a point $k_0$ of a single band
function $\lambda (k)$ such that $\lambda(0)=0$ starts with a
 nonzero
quadratic form\footnote{This form can be degenerate, as opposed to
our assumption in \cite[Corollary 11]{KuPin}.}, then the dimension
of the space of Floquet solutions of the equation $Pu=0$ on $X$ of
order at most $N$ with quasimomentum $k_0$ is equal to $h_{n,N}$,
i.e., it is equal to the dimension of the space of harmonic (in
the standard sense) polynomials of degree at most $N$ in
$\Real^n$. In particular, this condition is satisfied at
nondegenerate extrema of dispersion curves, i.e., the condition is
satisfied at nondegenerate spectral edges.
\end{enumerate}
In both cases an isomorphism can be provided explicitly as in the
previous theorem.
\end{corollary}

\mysection{Liouville type theorems for elliptic periodic
systems}\label{S:Liouv}

In this section we shall use the results of the previous sections to
establish Liouville theorems for periodic equations. We will
consider at the moment an arbitrary linear (square) matrix elliptic
operator $P$ on the space $X$ of the abelian covering $X\rightarrow
M$ with smooth $G$-periodic coefficients that satisfies the
assumptions made in Section \ref{S:notations}\footnote{The results
hold with the same proofs for periodic elliptic equations in
sections of periodic vector bundles on $X$.}. As before, without
loss of generality we can limit the consideration to the case $G
=\ZZ^n$. Any of the standard meanings of ellipticity of a system
would do, e.g., ellipticity in Petrovsky or Douglis--Nirenberg sense
\cite{Douglis}.

\begin{definition} \label{defLiouv} {\em Let $K\Subset X$ be a domain in $X$ such that
$\mathrel{\mathop{\cup }\limits_{g \in G}}gK=X$. For
$N=0,1,\ldots$, define the space

\bean \mathrm{V}_{N}(P)\!:=\!\left\{ u \mid Pu\!=\!0 \mbox{ in }
X, \mbox{ and } \sup_{g \in G} \!\left[\left|\left| u\right|
\right| _{L^2(gK)}\!\!(1\!+\!\left|g \right|)^{-N}\right] \!<\!
\infty \right\}. \eean

We say that {\em  the Liouville theorem of order $N\geq 0$ holds
true for the operator $P$} if $\dim \mathrm{V}_{N}(P)<\infty$.

We say that {\em the Liouville theorem holds true for the operator
$P$} if it holds for any order $N\geq 0$.}
\end{definition}

Abusing notations, we will call solutions in $\mathrm{V}_0(P)$
{\em bounded solutions}. Since obviously
$\mathrm{V}_{N_1}(P)\subset \mathrm{V}_{N_2}(P)$ for $N_1 < N_2$,
Liouville theorem of higher order implies the lower order ones. It
is not clear {\em a priori} that the converse holds. The next
results show that this is in fact true, at least in our situation
of abelian coverings. This observation apparently has failed to be
made in previous studies, which sometimes lead to investigations
of some individual Liouville theorems, e.g., for $N=0$ without
noticing their simultaneous validity for all $N$.

We will start with an auxiliary statement, which is an analog of
the classical theorem on the structure of distributions supported
at a single point. It is a generalization of Lemma 25 in
\cite{KuPin}, where Fredholm rather than semi-Fredholm property is
assumed. For the results of this section Lemma 25 in \cite{KuPin}
would be sufficient, while we need the full strength of the lemma
below to treat overdetermined systems and holomorphic functions in
Section \ref{S:overdet}.
\begin{lemma} \label{L:delta}
Let $T$ be a $C^\infty $-manifold and $P:T\rightarrow L(H_1,H_2)$
be a $C^\infty $-function with values in the space $L(H_1,H_2)$ of
bounded linear operators between the Hilbert spaces $H_1$ and
$H_2$. Assume that for each  $k\in T$ the operator $P(k)$ is right
semi-Fredholm (e.g., \cite{ZK}), i.e., it has a closed range and a
finite dimensional cokernel. Then

\begin{enumerate}
\item  If $P(k)$ is surjective for all points $k$ in $T$, then the
multiplication operator
\begin{equation}\label{E:surjective}
C^\infty (T,H_1)\stackrel{P(k)}{\rightarrow }C^\infty (T,H_2)
\end{equation}
is surjective.

\item For any fixed $k_0 \in T$ the dimension of the space of
functionals of the form
\begin{equation}\label{E:repr_single_k}
\left[ \sum\limits_{j\leq N} D_{j,k}(<g_{j},\phi >)\right] _{k_0}
\end{equation}
that are orthogonal to the range of the multiplication operator
(\ref{E:surjective}) is finite. Here $g_{j}$ are continuous linear
functionals on $H_2$, the pairing $<g_{j},\phi>$ denotes the
duality between $H_2^{*}$ and $H_2$, $D_{j,k}$ are linear
differential operators with respect to $k$ on $T$, and $N\geq 0$.

\item  If $P(k)$ is surjective for all points $k$ except of a finite
subset $F \subset T$, then any continuous linear functional $g$ on
the space of smooth vector functions $C^\infty (T,H_2)$ that
annihilates the range of the multiplication operator
\begin{equation}\label{E:mult_operator}
C^\infty (T,H_1)\stackrel{P(k)}{\rightarrow }C^\infty (T,H_2)
\end{equation}
has the form
\begin{equation} \label{E:repr}
<g,\phi >=\sum\limits_{k_l \in F}\left[ \sum\limits_{j\leq N}
D_{j,k}(<g_{j},\phi >)\right] _{k_l},
\end{equation}
in the notations of the previous statement of the lemma.

\end{enumerate}
\end{lemma}
\pf Let us establish the validity of the first statement of the
lemma. Due to the existence of partitions of unity, the statement
is local. Locally, following for instance the proof of Theorem 2.7
in \cite{ZK}, one can construct a smooth one-sided (right) inverse
$Q(k)$ to the operator function $P(k)$. Now multiplication by
$Q(k)$ provides a right inverse to (\ref{E:surjective}), which
proves the surjectivity.

To prove the second statement, let us consider a closed subspace
$M\subset H_1$ complementary to the kernel of $P(k_0)$. Then the
operator $P(k_0):M\to H_2$ is injective and Fredholm. This
injectivity property is preserved in a neighborhood $U$ of $k_0$. In
particular, the subspace $M(k)=P(k)(M)\subset H_2$ of finite
codimension forms a smooth subbundle $\mathcal{M}$ in $U\times H_2
\rightarrow U$. Now, any smooth $H_2$-valued function $f(k)$ such
that $f(k)\in M(k)$, belongs to the range of the operator
(\ref{E:surjective}). Hence, functionals orthogonal to the range can
be pushed down to the smooth sections of the finite dimensional
bundle $\mathcal{C}=\mathop{\cup}\limits_{k\in U} H_2/M(k)$ over
$U$. It is clear that the functional one gets on this bundle
preserves the structure (\ref{E:repr_single_k}). Such functionals on
a finite dimensional bundle, however, form a finite dimensional
space (for any fixed $N$).

The third statement can be proven analogously to the similar
statement in \cite[Corollary 1.7.2]{Ku}. Namely, under the
conditions of the statement, and taking into account the first claim
of the lemma, any functional annihilating the range of the operator
of multiplication by $P(k)$ must be supported at the finite set $F$
over which $P(k)$ is not surjective. We can reduce the consideration
to a neighborhood $U$ of a single point $k_0\in F$. Now, the proof
of the second statement reduces the functional to one supported at
the point $k_0$ and defined on smooth sections of a
finite-dimensional bundle $\mathcal{C}$ over $U$. Thus, the standard
representation of distributions supported at a point implies
(\ref{E:repr}). \qed

The next theorem shows that the existence of a polynomially
growing solution implies the existence of a nonzero bounded Bloch
solution (i.e., a solution automorphic with respect to a unitary
character of $G$).

\begin{theorem}\label{T:Bloch}
The equation $Pu=0$ has a nonzero polynomially growing solution if
and only if it has a nonzero Bloch solution with a real
quasimomentum (such a solution is automatically bounded), i.e., if
and only if the real Fermi surface $F_{P,\RR}=F_{P} \cap
\Real^{n}$ is not empty.
\end{theorem}
\pf Assume that $F_{P,\RR}=\emptyset$. Then $P(k)$ is surjective
for all $k\in \TT^n$. Indeed, if this were not the case, we could
find a nonzero functional of the type (\ref{E:functional}) with
$N=0$ and some $k$. According to the first statement of Lemma
\ref{L:Floq_struct}, this would mean the existence of a Bloch
solution. Now, the first statement of Lemma \ref{L:delta}
guarantees the surjectivity of the mapping
\[
C^\infty (\TT^n,{\mathcal E}^m)\stackrel{P(k)}{\rightarrow }C^\infty (\TT^n,{\mathcal %
E}^0)
\]
and hence the absence of any nontrivial functionals on $C^\infty
(\TT^n,{\mathcal E}^0)$ that annihilate the image of this mapping.
Since under the Floquet--Gelfand transform ${\mathcal U}$, any
polynomially growing solution $u(x)$ of $Pu=0$ is mapped to such a
functional, we conclude that $u=0$. \qed

The next result provides necessary and sufficient conditions under
which Liouville theorems hold for equations of the type we
consider. It also establishes that the validity of Liouville type
theorems does not depend on the order of polynomial growth.

\begin{theorem}
\label{T:Liouville} Under the conditions we have imposed on the
covering $X$ and the operator $P$, the following statements are
equivalent:
\begin{enumerate}

\item The number of points in the real Fermi surface $F_{P,\RR}$ is finite
(i.e., Bloch (or automorphic) solutions exist for only finitely
many unitary characters $\gamma_k$).

\item There exists $N\geq 0$ such that the Liouville
theorem of order $N$ holds true.

\item The Liouville theorem holds (i.e., it holds for any order $N$).
\end{enumerate}
\end{theorem}
\pf $(2)\Rightarrow (1)$. Any Bloch solution with a real
quasimomentum $k$ (i.e., corresponding to a unitary character) is
bounded and hence belongs to the space $\mathrm{V}_N(P)$ for any
$N$. Since such solutions with different characters are linearly
independent, the validity of the Liouville theorem for some value
of $N$ implies that the number of the corresponding characters is
finite. Since the characters of all Bloch solutions with real
quasimomenta constitute the real Fermi variety, $F_{P,\RR}$ is
finite.

The implication $(3)\Rightarrow (2)$ is obvious.

Let us now prove $(1)\Rightarrow (3)$. Let $u$ be a nonzero
polynomially growing solution. It can be interpreted as a
continuous functional on $\mathrm{C}^0$ annihilating the range of
the dual operator $P^*:\mathrm{C}^m\to \mathrm{C}^0$. After the
Floquet--Gelfand transform $\mathcal{U}$, one obtains a functional
on $C^\infty(\TT^n,\mathcal{E}^0)$ orthogonal to the range of the
operator
$$
C^\infty(\TT^n,\mathcal{E}^m) \stackrel{P(k)}{\rightarrow
}C^\infty(\TT^n,\mathcal{E}^0).
$$
By our assumption, $F_{P,\RR}$  is finite, thus the second and the
third statements of Lemma \ref{L:delta} and Theorem
\ref{T:Fl_dimen} finish the proof of statement $(3)$. \qed

While this theorem establishes conditions under which Liouville
theorem holds, it does not tell much about the dimensions of the
spaces $\mathrm{V}_N(P)$ of polynomially growing solutions,
besides those being finite. It also does not address the structure
of these solutions. The next result provides some estimates and
even explicit formulas for the dimensions, as well a
representation for the solutions.

\begin{theorem}\label{T:Liouville_dim} Suppose that the Liouville theorem holds
for an elliptic operator $P$ and let $d_N: =\dim \mathrm{V}_N(P)$.
Then the following statements hold:

\begin{enumerate}
\item Each solution $u \in \mathrm{V}_N(P)$ can be represented as a finite
sum of Floquet solutions:

\begin{equation}
u(x)= \sum\limits_{k \in F_{P,\RR}}\sum\limits_j u_{k,j}(x),
\label{li_repres}
\end{equation}
where each $u_{k,j}$ is a Floquet solution with a quasimomentum $k$,
and $F_{P,\RR}=F_P \cap \Real^n$.

\item For all $N\in \NN$, we have
$$
d_N\leq d_0 q_{n,N}<\infty\,,
$$
where $q_{n,N}$ is the dimension of the space of all
polynomials of degree at most $N$ in $n$ variables.

\item Assume that the spectra of the operators $P(k)$ are discrete for all $k$
and that for each real quasimomentum $k\in F_{P,\RR}$ the
conditions of Theorem \ref{T:Fl_dimen} are satisfied. Then for
each $N\geq 0$ the dimension $d_N$ of the space $\mathrm{V}_N(P)$
is equal to
\begin{equation}\label{E:dimension}
   \sum\limits_{k\in F_{P,\RR}}
   r_k\left[  \left(%
\begin{array}{c}
  n+N \\
  N\\
\end{array}%
\right) - \left(%
\begin{array}{c}
  n+N-l_0(k) \\
  N-l_0(k)\\
\end{array}%
\right) \right].
\end{equation}
Here $r_k$ and $l_0(k)$ are respectively the multiplicity of the
zero eigenvalue and the order of the first nonzero Taylor term of
the dispersion relation at the point $k \in F_{P,\RR}$ (see these
notions explained before Theorem \ref{T:Fl_dimen}). The terms in
this sum are the dimensions of the spaces of
$\lambda_{l_0(k)}(D)$-harmonic polynomials, and polynomially
growing solutions can be described in terms of these polynomials
analogously to Theorem \ref{T:Fl_dimen}.
\end{enumerate}
\end{theorem}

\pf All these statements follow immediately from Lemma
\ref{L:Floq_struct}, Theorem \ref{T:Fl_dimen}, Lemma
\ref{L:delta}, and Theorem \ref{T:Liouville}. \qed

\mysection{Examples of Liouville theorems for specific
operators}\label{S:examples}

Let us recall that the $L^2$-spectrum of the operator $P$ is the
union over $k\in \mathrm{B}$ of the spectra of $P(k)$ (e.g.,
\cite[Theorem 4.5.1]{Ku} and \cite[Theorem XIII.85]{RS}). In other
words, the spectrum of $P$ coincides with the range of the
dispersion relation over the Brillouin zone $\mathrm{B}$. We have
also discussed that the real Fermi surface for $P$ is just the
zero level set for the dispersion relation over the Brillouin
zone. Theorem \ref{T:Liouville} of the previous section shows that
the Liouville theorem holds if and only if this Fermi surface is
finite\footnote{In particular, outside the spectrum the Liouville
theorem holds vacuously, according to Theorem \ref{T:Liouville}
and Theorem 4.5.1 in \cite{Ku}.}. One expects this to happen
normally at `extrema' of the dispersion relation (albeit this is
neither necessary, nor sufficient). In other words, speaking for
instance about the {\em selfadjoint case}, one should expect
Liouville theorem to hold mainly when zero is at the edge of a
spectral gap, although it is possible in principle to have
interior points of the spectrum where such a thing could occur as
well\footnote{Two spectral zones with touching edges could provide
such an example.}. One notices that the cases considered in
\cite{AL,LW1,MS} all correspond to zero being at the bottom of the
spectrum (in the second-order nonselfadjoint case we mean by the
bottom of the spectrum the generalized principal eigenvalue, see
\cite{Agmon_positive,KuPin,LP} and Section 5.4 below). This
explains why the homogenization techniques employed in these
papers could be successful. Indeed, homogenization works exactly
at the bottom of the spectrum. The results of this work show that
one should also consider internal spectral edges, where the
standard homogenization does not apply.

So, let us now try to look at some examples where one can apply
the results of the previous sections. Theorem \ref{T:Liouville} is
applicable to any elliptic periodic equation or system of
equations on any abelian covering of a compact manifold. According
to this theorem, one only needs to establish the finiteness of the
real Fermi surface. On the other hand, Theorem
\ref{T:Liouville_dim}, which provides formulas for the dimensions
of the spaces of polynomially growing solutions, as well as some
representations of such solutions, is more demanding. It requires
that one guarantees discreteness of the spectra of the `cell'
operators $P(k)$, and most of all, requires understanding of the
analytic structure of the dispersion curve near the Fermi surface.
Concerning this structure, it is expected that the following is
true:
\begin{conjecture}\label{C:conj}
Let $P$ be a `generic' selfadjoint second-order elliptic operator
with periodic coefficients, and let $(\lambda_{-},\lambda_{+})$ be
a nontrivial gap in its spectrum. Then each of the gap's endpoints
is a unique (modulo the dual lattice) and nondegenerate extremum
of a single band function $\lambda_{j}(k)$.
\end{conjecture}

This conjecture is crucial in many problems of mathematical physics,
spectral theory, and homogenization (see its further discussion in
Section \ref{S:remarks}) and is widely believed to hold.
Unfortunately, the only known theorem of this kind is the recent
result of \cite{KloppRalston}, which states that generically a gap
edge is an extremum of a single band function. A similar conjecture
probably holds for equations on abelian coverings of compact
manifolds. Theorem \ref{T:Liouville} shows that validity of
Conjecture \ref{C:conj} would imply the following statement:

\begin{theorem}\label{T:conditional} If Conjecture
\ref{C:conj} holds true, then `generically', at the spectral edges
of a selfadjoint elliptic operator of second-order with periodic
coefficients on $\RR^n$, the Liouville theorem holds, and for each
$N\geq 0$ the dimension of the space $\mathrm{V}_{N}$ is equal to
$h_{n,N}$, the dimension of the space of all harmonic polynomials
of order at most $N$ in $n$ variables.
\end{theorem}

In the similar situation on abelian coverings of compact
manifolds, the rank of the deck group would appear in the answer
rather than the dimension of the manifold (see Section
\ref{S:remarks}).

At the bottom of the spectrum, however, much more is known
\cite{BS1,BS3,BS_intern, FKT, Kirsch, Pr, Shter}. Even this
limited (in terms of spectral location) information together with
theorems \ref{T:Liouville} and \ref{T:Liouville_dim} provides one
with many specific examples that go far beyond the equations
considered in \cite{AL,LW1,MS}.

The first trivial remark is that if zero is outside the spectrum
of the operator $P$, then according to Theorem \ref{T:Liouville}
the Liouville property holds vacuously. Indeed, in this case the
real Fermi surface is empty, and hence the equation $Pu=0$ has no
polynomially growing solutions. Let us look now at some less
trivial examples.

\subsection{Schr\"{o}dinger operators}

Let $X\to M$ be, as before, a noncompact abelian covering of a
$d$-dimensional compact manifold and $H=-\Delta + V(x)$ be a
Schr\"{o}dinger operator on $X$ with a periodic real valued
potential $V \in L^{r/2}_{\mathrm{loc}}(X),\,r>d$. Then the result
of \cite{Kirsch} for $X=\RR^d$ and of \cite{KobSun} in the general
case states that the lowest band function $\lambda_1(k)$ has a
unique nondegenerate minimum $\Lambda_{0}$ at $k=0$. All other
band functions are strictly greater than $\Lambda_{0}$. In
particular, the bottom of the spectrum of the operator $H$ is at
$\Lambda_{0}$. Let us assume that $\Lambda_{0}=0$ (or replace the
operator by $H=-\Delta + V(x)-\Lambda_{0}$). Then theorems
\ref{T:Liouville} and \ref{T:Liouville_dim} become applicable,
since the real Fermi surface consists of a single point $k=0$ and
the band function has a simple nondegenerate minimum at this point
(i.e., $r=1$ and $l_0=2$ in the notations of Theorem
\ref{T:Liouville_dim}). Thus, one obtains
\begin{theorem}\label{T:schrod}
\begin{enumerate}
\item The Liouville theorem holds for the operator $H-\Lambda_{0}$.
\item The dimension of the space $\mathrm{V}_N (H-\Lambda_{0})$  equals $h_{n,N}$
(where $n$ is the free rank of $G$).
\item Every solution $u\in \mathrm{V}_N(H-\Lambda_{0})$ of the equation
$Hu-\Lambda_{0}u=0$ is a Floquet solution of order $N$ with
quasimomentum $k=0$, i.e., it can be represented as
$$
u(x)=\sum\limits_{|j|\leq N} [x]^j p_j(x)
$$
with $G$-periodic functions $p_j$.
\end{enumerate}
\end{theorem}

\subsection{Magnetic Schr\"{o}dinger operators}

Although the first result of this subsection holds also in the
general situation of abelian coverings, we will present it for
simplicity in the flat case.

Consider the following selfadjoint magnetic Schr\"{o}dinger
operator on $\RR^n$
\begin{equation}\label{E:magnetic}
H=\left[\mathrm{i} \nabla +A(x)\right]^2+V(x)
\end{equation}
with periodic electric and magnetic potentials $V$ and $A$,
respectively\footnote{One can see that the considerations and the
result of this subsection immediately extend to more general
abelian coverings. We avoid doing so, in order not to introduce
any new notions required for defining the magnetic operators in
such a setting.}. The introduction of a periodic magnetic
potential into the operator is known to change properties of the
Fermi surface significantly (e.g., \cite{FKT,Shter}). However,
small magnetic potentials do not destroy the properties of our
current interest. Indeed, let $V$ be as in the previous statement,
then there exists $\varepsilon>0$ (depending on $V$) such that for
any periodic real valued magnetic potential $A$ such that
$$ \left|\left| A
\right|\right|_{L^{r}(\mathbb{T}^{n})} < \varepsilon $$ and
\begin{equation}\label{A0}
\int\limits_{\mathbb{T}^n}A(x)\,\mathrm{d}x=0 \end{equation} the
following holds true: The lowest band function $\lambda_1(k)$ of
$H$ attains a unique nondegenerate minimum $\Lambda_{k_0}$ at a
point $k_{0}$ (albeit, not necessarily at $k=0$). All other band
functions are strictly greater than $\Lambda_{k_0}$. Indeed, when
both $V$ and $A$ are sufficiently small, this is proven in
\cite{FKT}. It is not hard, though, to allow for arbitrary
electric and small magnetic potential. Indeed, the case when $A=0$
is covered by the result of \cite{Kirsch} (see the previous
subsection). Now, the statement of Lemma \ref{L:Fermi} (see also
\cite[Theorem 4.4.2]{Ku}) can be easily extended without any
change in the proof to include analyticity with respect to the
potentials (e.g., \cite{FKT}). Namely, there exists an entire
function $f(k,\lambda,A,V)$ of all its arguments such that
$f(k,\lambda,A,V)=0$ is equivalent to
 $$ (k,\lambda)\in B_{(\mathrm{i}\nabla +A)^2+V}\,, $$
where as before $B_H$ is the Bloch variety of the operator $H$.
This, together with the just mentioned result of \cite{Kirsch} for
$A=0$, imply that the lowest band function $\lambda_1(k)$ of $H$
attains a unique nondegenerate minimum $\Lambda_{k_0}$ at a point
$k_{0}$ for sufficiently small magnetic potentials\footnote{The
degree of smallness of $A$ depends on the electric potential
$V$.}.

Now one uses theorems \ref{T:Liouville} and \ref{T:Liouville_dim}
again to obtain

\begin{theorem}\label{T:magnetic} Assume that  $V$, $A$, and $k_0$
satisfy the above assumptions, then
\begin{enumerate}
\item The Liouville theorem holds for $u\in \mathrm{V}_N(H-\Lambda_{k_0})$.
\item Any solution $u\in \mathrm{V}_N(H-\Lambda_{k_0})$ is representable in
the Floquet form
$$
v(x)=\mathrm{e}^{\mathrm{i}k_{0}\cdot x}\sum\limits_{|j|\leq
N}x^jp_j(x)
$$
with periodic functions $p_j(x)$.
\item The dimension of the space $\mathrm{V}_N(H-\Lambda_{k_0})$ is
equal to $h_{n,N}$.
\end{enumerate}
\end{theorem}

Note that the normalization (\ref{A0}) can be always achieved by a
gauge transformation, which does not affect the spectrum and the
Liouville property.

Besides magnetic Schr\"{o}dinger operators with small magnetic
potentials, another very special subclass of periodic magnetic
Schr\"{o}dinger operators is formed by the so called {\em Pauli
operators}. Consider the following Pauli operators in $\RR^2$:
\begin{equation}\label{E:Pauli}
P_\pm=(\mathrm{i}\nabla +A)^2 \pm
(\partial_{x_1}A_2-\partial_{x_2}A_1),
\end{equation}
where $A(x)=(A_1(x),A_2(x))$ is periodic. The structure of the
dispersion curves at the bottom of the spectrum of such an
operator was studied in \cite{BS1}. It was shown that the
dispersion relation for $P_\pm$ attains at the point $k=0$ its
single nondegenerate minimum with the value $\Lambda_0=0$. This
implies that the result for the Pauli operators holds exactly like
for the Schr\"{o}dinger operator in the previous subsection, and
in fact in a more precise form, since the minimal value
$\Lambda_0$ of the band function is known. Namely, every solution
$u\in \mathrm{V}_N(P_\pm)$ of the equation $P_\pm u=0$ on
$\mathbb{R}^2$ is representable in the form (\ref{polyn}). The
dimension of the space $\mathrm{V}_N(P_\pm)$ is equal to
$h_{2,N}=2N+1$, which is the dimension of the space of all
harmonic polynomials of order at most $N$ in two variables.

The cases above of small magnetic potentials and of Pauli
operators form a rather special subclass of periodic magnetic
Schr\"{o}dinger operators, in the sense that one still finds a
single nondegenerate minimum of band functions at the bottom of
the spectrum. However, this is not true anymore for the whole
class of periodic magnetic Schr\"{o}dinger operators, which should
influence significantly our Liouville theorems. Although there is
not much known here yet, the results of \cite{Shter} provide the
first glimpse at the possibilities. In that paper the following
operator in $\RR^2$ is considered:
$$
M_t=(\mathrm{i}\nabla +tA(x))^2 -t^2,
$$
where $t\in\RR$ and the magnetic potential $A(x)$ is $(0,a(x_1))$
with the $1$-periodic function $a(x_1)$ that is equal to
$\,\mathrm{sign} (x_1)$ on $[-0.5,0.5]$. The main result of
\cite{Shter} is that for $|t|<2\sqrt{3}$ the bottom of the
spectrum of $M_t$ is $0$ and it is attained at the quasimomentum
$k=0$, where the band function has a simple nondegenerate minimum.
This implies immediately the same Liouville theorem as for the
Pauli operators. The situation changes for $|t|=2\sqrt{3}$, when
this minimum (which is still $0$ and is attained at $k=0$) becomes
degenerate. Namely, it is shown in \cite{Shter} that the lowest
eigenvalue $\lambda_1(k)$ of the operators $M_t(k)$ for
$|t|=2\sqrt{3}$ has the following Taylor expansion at $k=0$:
$$
\lambda_1(k)=k_1^2+\frac{1}{42}k_2^4-\frac{1}{10}k_1^2k_2^2+O(|k|^6).
$$
This shows the degeneration of the Hessian. On the other hand,
since the first nonzero homogeneous term of the expansion is of
order two, Theorem \ref{T:Liouville_dim} still says that the
Liouville theorem sounds the same as before (i.e., $\dim
\mathrm{V}_N(M_{\pm2\sqrt{3}})=2N+1$). This changes, though, for a
little bit larger values of $|t|$, when the bottom of the spectrum
shifts to the negative half-axis and two points of extremum appear
with a nonvanishing Hessian at these quasimomenta. This implies
that the Liouville theorem will apply after shifting to the bottom
of the spectrum (i.e., after adding an appropriate positive scalar
term to $M_t$), but the dimensions of the spaces $\mathrm{V}_N$ of
polynomially growing solutions will increase above $h_{2,N}$.

\begin{theorem}\label{T:pauli}
\begin{enumerate}
\item The Liouville theorem holds for the operators $P_\pm$ and $M_t$
for $|t|\leq 2\sqrt{3}$.
\item The dimensions of the corresponding spaces $\mathrm{V}_N$ are
equal to $h_{2,N}=2N+1$.
\item Every solution $u\in \mathrm{V}_N$ can be
represented as
$$
u(x)=\sum\limits_{|j|\leq N} x^j p_j(x)
$$
with $G$-periodic functions $p_j$.
\item For a sufficiently small $\varepsilon >0$ and the values
$|t|\in (2\sqrt{3},2\sqrt{3}+\varepsilon)$, the Liouville theorem
holds at the bottom of the spectrum of the operator $M_t$, while
the dimension of the space $\mathrm{V}_N(M_t)$ is equal to
$2h_{2,N}$.
\end{enumerate}
\end{theorem}

It is also conjectured in \cite{Shter} that for some values of
$t$, the operator $M_t$ with the magnetic potential equal to
$A(x)=(\mathrm{sign}(x_2),\mathrm{sign}(x_1))$ on the cube
$[-0.5,0.5]^2$ and $\ZZ^2$-periodic, will exhibit a complete
degeneration of the bottom of the spectrum in the sense of
complete vanishing of the Hessian. If this happens to be true,
then according to Theorem \ref{T:Liouville_dim}, it would lead to
an increase in the dimensions of the spaces $\mathrm{V}_N$.

\subsection{Operators admitting regular factorization}

A very general and wide class of selfadjoint second-order periodic
operators $H$ on $\RR^n$, for which one can study thoroughly the
dispersion curves at the bottom of the spectrum, was introduced in
\cite{BS3,BS2}. It consists of operators that allow for what the
authors of \cite{BS3,BS2} call {\em regular factorization}:
\begin{equation}\label{E:factor}
H=\overline{f(x)}b(D)^*G(x)b(D)f(x).
\end{equation}
Here $b(D)=\sum b_j D_{x_j}$ is a constant coefficients
first-order linear differential operator, the periodic function
$f(x)$ is such that $f,f^{-1}\in L^\infty(\RR^n)$, and the
periodic matrix-function $G(x)$ satisfies $c_0 I \leq G(x) \leq
c_1 I$ for some $0<c_0\leq c_1$. Then, it was shown in \cite{BS2}
that the bottom of the spectrum of $H$ is $0$ and it is attained
at the quasimomentum $k=0$, where the band function has a simple
nondegenerate minimum. As in the previous subsections, this
implies

\begin{theorem}\label{T:factor}
\begin{enumerate}
\item The Liouville theorem holds for the operator $H$
(\ref{E:factor}).
\item $\dim \mathrm{V}_N (H)=h_{n,N}$.
\item Every solution $u\in \mathrm{V}_N(H)$ can be
represented as
$$
u(x)=\sum\limits_{|j|\leq N} x^j p_j(x)
$$
with $G$-periodic functions $p_j$.
\end{enumerate}
\end{theorem}

\subsection{Nonselfadjoint second-order operators}

Throughout this subsection, we again consider the flat case
$X=\RR^n$ only. However, the result still holds on general abelian
coverings, which does not require any significant change in the
proof.

In all the examples that were discussed so far in this section, we
considered only selfadjoint operators. There is, however, a class
of (in general nonselfadjoint) second-order operators in $\RR^n$,
which can be studied thoroughly and which plays an important role
in probability theory.

Consider second-order uniformly elliptic operators on
$\mathbb{R}^n$ of the following form (\ref{E:operator}):
\begin{equation}\label{E:operator2}
L=-\sum_{i,j=1}^n a_{ij}(x)\partial _{i}\partial _{j}+\sum_{i=1}^n
b_i(x)\partial _i+c(x) \qquad x\in \mathbb{R}^n
\end{equation}
with coefficients that are {\it real} and periodic.

For an operator of this type, an important function $\Lambda (\xi
):\RR^n\rightarrow \RR$ can be introduced\footnote{The notation
$\Lambda$ has been used before in this paper to denote a somewhat
different -albeit related - object. However, these two notations
do not collide in this subsection, so the reader should not get
confused.}, whose properties were studied in detail in
\cite{Agmon_positive,LP,Pr}. It is defined by the condition that
the equation
\[
Lu=\Lambda (\xi )u
\]
has a {\em positive} Bloch solution of the form
\begin{equation}\label{positiveBloch}
u_{\,\xi }(x)=\mathrm{e}^{\xi \cdot x}p_{\,\xi }(x),
\end{equation}
where $p_{\,\xi }(x)$ is $G$-periodic.

We also need to define the following number:
\begin{equation}
\Lambda_{0} =\max_{\xi \in \RR^n}\Lambda (\xi ).  \label{Lambda}
\end{equation}
It follows from \cite{Agmon_positive,LP} that $\Lambda_{0}$ can
also be described as follows:
\begin{equation}\label{genev}
\Lambda_0= \sup\{\gl \in \Real \; |\; \exists u>0 \mbox{ such that
} (L-\gl)u= 0 \mbox{ in } \Real^n\}.
\end{equation}
In the selfadjoint case, $\Lambda_{0}$ is the bottom of the
spectrum of the operator $L$. The common name for $\Lambda_{0}$ is
{\em the generalized principal eigenvalue} of the operator $L$ in
$\Real^n$.

We assemble the information we need about the function
$\Lambda(\xi)$ and the generalized principal eigenvalue
$\Lambda_{0}$ in the following lemma. The reader can find proofs
of these statements in \cite{Agmon_positive,LP,Pr} and more
detailed references in \cite{KuPin}.

\begin{lemma}\label{Lambda-lemma}
\begin{enumerate}
\item  The value $\Lambda (\xi )$ is uniquely determined for any $\xi \in
\Real^n$.

\item  The function $\Lambda (\xi )$ is bounded from above, strictly
concave, analytic, and has a nonzero gradient at all points except
at its maximum point.

\item  Consider the operator
$$L(\xi )=\mathrm{e}^{-\xi\cdot x}L\mathrm{e}^{\xi\cdot x}=L(x,D-i\xi) $$ on the
torus $\mathbb{T}^n$.  Then $\Lambda (\xi )$ is the principal
eigenvalue of $L(\xi )$ with  a positive eigenfunction $p_{\,\xi
}$. Moreover, $\Lambda (\xi )$ is algebraically simple.

\item The Hessian of $\Lambda (\xi )$ is nondegenerate at all points.

\item $\Lambda _{0}\geq 0$ if and only if the operator $L$ admits a positive
(super-) solution. This condition is satisfied in particular when
$c(x)\geq 0$.

\item $\Lambda _{0}\geq 0$ if and only if the operator $L$ admits a positive
solution of the form (\ref{positiveBloch}).

\item $\Lambda _{0}=0$ if and only if the equation $Lu=0$ admits
exactly one normalized positive solution in $\Real^n$.

\item If $c(x)=0$, then $\Lambda_{0} =0$ if and only if
$\int\limits_{{\mathbb T}^n}b(x)\psi (x)\,\mathrm{d}x=0$, where
$\psi $ is the principal eigenfunction of $L^{*}$ on ${\mathbb
T}^n$, and $b(x)=(b_1(x),\ldots,b_n(x))$. In particular,
divergence form operators satisfy this condition.

\item Let $\xi\in \Real^n$, and assume that  $u_\xi(x)=\mathrm{e}^{\xi \cdot x}p_\xi (x)$
and $u^*_{-\xi}$ are positive Bloch solutions of the equations
$Lu=0$ and $L^*u=0$, respectively. Denote by $\psi$ the periodic
function $u_\xi u^*_{-\xi}\,$. Consider the function
\[
\tilde{b}_i(x)= b_i(x)-2\sum_{j=1}^n a_{ij}(x)\left\{\xi_j
+\left[p_\xi(x)\right]^{-1}\partial_{j}p_\xi(x)\right\}.
\]
Then $\Lambda_{0} =0$ if and only if
\[
\int\limits_{\mathbb{T}^n}\tilde{b}_i(x)\psi
(x)\,\mathrm{d}x=0\qquad i=1,\ldots,n.
\]
\end{enumerate}
\end{lemma}
Now one can describe Liouville type theorems for an operator of
the form (\ref{E:operator2}) assuming that $\Lambda_{0} \geq 0$.
This assumption implies that the operator admits a positive
supersolution.

It is shown in \cite{KuPin} that if $\Lambda(0)\geq 0$, then the
Fermi surface $F_{L}$ can touch the real space only at the origin
(modulo the reciprocal lattice $G ^{*}=\left( 2\pi \ZZ\right) ^n$)
and in this case $\Lambda(0)= 0$.

In fact, we have the following stronger result which extends Lemma
15 in \cite{KuPin} to nonreal $\lambda$. In addition, the proof of
the statement below is more elementary than in that lemma.
\begin{lemma}\label{L:lemext}
 Let $k=\gamma-i \xi\in \mathbb{C}^n$. If $Re \, \lambda < \Lambda(\xi)$,
 then
$(k,\lambda)$ does not belong to the Bloch variety $B_L$ of the
operator $L$. Moreover, if $Re\, \lambda = \Lambda(\xi)$, then
$(k,\lambda) \in B_L$ if and only if $\gamma\in G^*$, $\xi$
belongs to the zero level set $\Xi$ of $\Lambda(\xi)$, and $Im\,
\lambda =0$.
\end{lemma}
\pf Without loss of generality, we may assume that $\xi=0$,
$\Lambda(\xi)=0$, and $L \mathbf{1}=0$. Assume that $(k,\lambda)
\in B_L$, and let $u(x)=\mathrm{e}^{\mathrm{i}\gamma\cdot x}p(x)$
be a Floquet solution with a quasimomentum $k$ of the equation
$$Lu=\lambda u \qquad \mbox{in } \mathbb{R}^n,$$ where $Re\,
\lambda\leq 0$. Take the complex conjugate
$$L\bar{u}=\bar{\lambda} \bar{u} \qquad
\mbox{in } \mathbb{R}^n ,$$ and compute $$L(|u|^2)=
\bar{u}Lu+uL\bar{u} -2\sum_{i,j=1}^n
a_{ij}u_{x_i}\bar{u}_{x_j}=2Re\, \lambda |u|^2-2\sum_{i,j=1}^n
a_{ij}u_{x_i}\bar{u}_{x_j}.$$ Notice that for each $\zeta\in
\mathbb{C}^n$, we have
$$\sum_{i,j=1}^n a_{ij}\zeta_i\bar{\zeta}_{j}=\sum_{i,j=1}^n
a_{ij}Re\, \zeta_iRe\, {\zeta}_{j}+\sum_{i,j=1}^n a_{ij}Im\,
\zeta_iIm\, {\zeta}_{j}\geq 0.$$ Therefore,
\begin{equation}\label{eqlusq}
L(|u|^2)\leq  0.
\end{equation}
 Thus, $|u|^2=|p(x)|^2$ is a periodic nonnegative subsolution of the
equation $Lu=0$ in $\mathbb{R}^n$. By the strong maximum principle
$|u(x)|^2=\mbox{constant}$. In particular, $L(|u|^2)=0$. Since we
have equality in (\ref{eqlusq}), it follows that $Re\, \lambda=0$
and
$$\sum_{i,j=1}^n a_{ij}u_{x_i}\bar{u}_{x_j}=
\sum_{i,j=1}^n a_{ij}Re\, u_{x_i}Re\, u_{x_j}+\sum_{i,j=1}^n
a_{ij}Im\, u_{x_i}Im\, u_{x_j}=0.$$ It follows that
$u=\mbox{constant}$ and therefore,
$\mathrm{e}^{\mathrm{i}\gamma\cdot x}$ is a periodic function.
Consequently, $\gamma\in G^*$. Since $L \mathbf{1}=0$, it follows
that $Im\, \lambda=0$.
 \qed

Lemma \ref{L:lemext} and Theorem \ref{T:Liouville} imply that if
$\Lambda(0)>0$, then the Liouville theorem holds vacuously, and
the equation $Lu=0$ does not admit any nontrivial polynomially
growing solution.

Suppose now that $\Lambda(0)= 0$ (i.e., $0\in F_{L,\RR}$). Then by
Lemma \ref{L:lemext} $F_{L,\RR}=\{0\}$. Moreover, Lemma
\ref{Lambda-lemma} implies that if $\Lambda_{0}>0$, then the point
$k=0$ is a noncritical point of the dispersion relation, and if
$\Lambda_{0}=0$ then $k=0$ is a nondegenerate extremum. In terms
of Theorem \ref{T:Liouville_dim}, we have in the first case
$l_0=1$, while in the second $l_0=2$. Moreover, in both cases
$r=1$. Theorems \ref{T:Liouville} and \ref{T:Liouville_dim} imply
now that the Liouville theorem holds, and every solution $u\in
\mathrm{V}_N(L)$ is representable in the form (\ref{polyn}). The
dimension of the space $\mathrm{V}_N(L)$ is equal to $h_{n,N}$ in
the case when $\Lambda_{0}=0$, and to $q_{n-1,N}$ when
$\Lambda_{0}>0$.

We summarize these results in the following statement:
\begin{theorem}\label{T:non-self}Let $L$ be a periodic operator
of the form (\ref{E:operator2}) such that $\Lambda_{0} \geq 0$.
Then
\begin{enumerate}

\item The Liouville theorem holds vacuously if $\Lambda(0)> 0$, i.e., the equation
$Lu=0$ does not admit any nontrivial polynomially growing
solution.

\item If $\Lambda(0)= 0$ and $\Lambda_{0}>0$, then the Liouville theorem holds for
$L$,
$$
\dim \mathrm{V}_N
(L)=q_{n-1,N}=\left(\begin{array}{c}n+N-1\\N\end{array}\right),
$$
and every solution $u\in \mathrm{V}_N(L)$ of the equation $Lu=0$
can be represented as
$$
u(x)=\sum\limits_{|j|\leq N} x^j p_j(x)
$$
with $G$-periodic functions $p_j$.
\item If $\Lambda(0)= 0$ and $\Lambda_{0}=0$, then the Liouville theorem holds for
$L$,
$$
\dim \mathrm{V}_N (L)=h_{n,N}=
\left(\begin{array}{c}n+N\\N\end{array}\right)-\left(\begin{array}{c}n+N-2\\N-2\end{array}\right),
$$
and every solution $u\in \mathrm{V}_N(L)$ of the equation $Lu=0$
can be represented as
$$
u(x)=\sum\limits_{|j|\leq N} x^j p_j(x)
$$
with $G$-periodic functions $p_j$.
\end{enumerate}
\end{theorem}

It is interesting to notice that for selfadjoint operators, one
never faces the case $l=1$ at the spectral edges, as it does
happen above when $\Lambda(0)= 0$ and $\Lambda_{0}>0$.

\mysection{Liouville theorems for overdetermined elliptic periodic
systems and for holomorphic functions}\label{S:overdet}

We will show here how the techniques and results of the preceding
sections can be applied to some overdetermined elliptic systems, in
particular to the Cauchy--Riemann operators on abelian covers of
compact complex manifolds. The main construction stays the same, so
we will be brief in its description.

We consider as before an abelian covering $X\stackrel{G}\mapsto M$
of a compact Riemannian manifold. Let also $L_j$ ($j=0,1,\ldots$)
be finite dimensional smooth vector bundles on $M$ equipped with a
Hermitian metric and $\mathcal{L}_j$ be the sheaves of their
smooth sections. Suppose that we have an elliptic complex of
differential operators
$$
    0\rightarrow \mathcal{L}_0 \stackrel{P}{\rightarrow}
    \mathcal{L}_1 \stackrel{P_1}{\rightarrow}
    \mathcal{L}_2 \stackrel{P_2}{\rightarrow}\ldots.
$$
We can lift this complex to an elliptic complex on $X$ for which we
will use the same notations:
\begin{equation}\label{E:Mcomplex}
    0\rightarrow \mathcal{L}_0 \stackrel{P}{\rightarrow}
    \mathcal{L}_1 \stackrel{P_1}{\rightarrow}
    \mathcal{L}_2 \stackrel{P_2}{\rightarrow}\ldots.
\end{equation}
Notice that the lifted bundles have a natural $G$-action with respect
to which the operators are periodic.

We will be interested in the spaces $\mathrm{V}_N(P)$ of the
global solutions $u(x)$ of the equation $Pu=0$ on $X$ that have
polynomial growth in the sense that for any compact $K \in X$
\begin{equation}\label{growth_overdet}
\|u\|_{L^2(gK,L_0)} \leq C (1+|g|)^N \qquad \forall g\in G
\end{equation}
for some $C$ and $N$. As before, without loss of generality we can
assume that $G=\ZZ^n$. The Floquet--Gelfand transform reduces the
elliptic complex (\ref{E:Mcomplex}) and its dual to the direct
integrals with respect to the quasimomentum $k$ of elliptic
complexes on $M$
\begin{equation}\label{E:complex_Floquet}
    0\rightarrow \mathcal{E}^1 \stackrel{P(k)}{\rightarrow}
    \mathcal{E}^2 \stackrel{P_1(k)}{\rightarrow}\ldots,
\end{equation}
and
\begin{equation}\label{E:complex_dual}
    0\leftarrow \mathcal{F}^1 \stackrel{P^*(k)}{\leftarrow}
    \mathcal{F}^2 \stackrel{P_1^*(k)}{\leftarrow}\ldots,
\end{equation}
where $\mathcal{E}^j, \mathcal{F}^j$ are appropriately defined
analytic Banach vector bundles over $\CC^n$ (a more detailed setup
of the relevant spaces can be found for instance in \cite{Pa2}.).
As previously, solutions of polynomial growth of $Pu=0$ after
applying the Floquet--Gelfand transform become functionals on
$C^\infty(\TT^n,\mathcal{E}^1)$ orthogonal to the range of the
operator
\begin{equation}\label{E:complex_orthog}
C^\infty(\TT^n,\mathcal{F}^2) \stackrel{P^*(k)}{\rightarrow}
    C^\infty(\TT^n,\mathcal{F}^1).
\end{equation}
Notice that due to the ellipticity of the complex and the
compactness of the base, it is Fredholm as a complex of operators
between the appropriate Sobolev spaces (e.g., \cite[Vol. III,
Sections 19.4 and 19.5]{Horm_differ}, \cite[Section 3.2., in
particular Theorem 13]{Rempel}, \cite[Section IV.5]{Wells}).
Hence, the operators $P^*(k)$ on $M$ belong to the set $\Phi_r$ of
the right semi-Fredholm operators between the corresponding
Sobolev spaces of sections. This means that they have closed
ranges of finite codimension, while having infinite dimensional
kernels. It is also clear that as in the elliptic case before,
$P(k)$ depends analytically (in fact, polynomially) on $k$. The
Fermi surface $F_P$ of $P$ is introduced as the set of all values
of $k$ for which the equation $P(k)u=0$ has a nonzero
solution\footnote{It can be shown similarly to the elliptic case
that the Fermi surface, as in the elliptic case, is an analytic
subset of $\CC^n$.}. As before, it coincides with the set where
the dual operator $P^*(k)$ has a nonzero cokernel. This enables us
to carry over the considerations of the previous section to prove
the following result:
\begin{theorem}\label{T:overdet}
The following conditions are equivalent:
\begin{enumerate}
\item The (real) Fermi surface $F_{P,\RR}$ is finite.

\item For some $N\geq 0$ the dimension $d_N$ of the space
$\mathrm{V}_N (P)$ of
solutions of polynomial growth of order $N$ is finite (i.e., the
Liouville theorem of order $N$ holds).

\item The dimension of $\mathrm{V}_N (P)$ is finite for any $N\geq 0$
(i.e., the Liouville theorem holds) and $d_n\leq d_0 q_{n,N}$.
\end{enumerate}
If one of the above conditions holds, then any solution of
polynomial growth is a linear combination of Floquet solutions.
\end{theorem}
\pf Since any point $k \in F_{P,\RR}$ provides a bounded Bloch
solution and those for different values of $k$ (modulo the dual
lattice) are linearly independent, the implications (3)
$\Rightarrow$ (2) $\Rightarrow$ (1) are obvious. Let us now assume
(1) and prove (3). According to Lemma \ref{L:delta}, all
functionals of order $N$  orthogonal to the range of
(\ref{E:complex_orthog}) can be expressed as linear combinations
over $k\in F_{P,\RR}$ of the form (\ref{E:repr}). The same Lemma
together with Lemma \ref{L:Floq_struct} and the finiteness of
$F_{P,\RR}$ show the finite dimensionality of this space and the
Floquet representation for all solutions of this type. \qed

We apply the above theorem to the following interesting case.
\begin{theorem}\label{T:holom}
Let $X \stackrel{G}{\rightarrow} M$ be an abelian cover of a
compact (complex) analytic manifold $M$. Let $X$ be equipped with
the lift of an arbitrary Riemannian metric from $M$ and $\rho$ be
the corresponding distance. For $N= 0, 1,\ldots $, let  $A_N (X)$
be the space of all analytic functions on $X$ growing not faster
than $C(1+\rho (x))^N$. Then:
\begin{enumerate}
\item The real Fermi surface of the $\bar{\partial}$-operator
defined on functions on $X$ contains only the origin:
$F_{\bar{\partial},\RR}=\{0\}$.

\item $A_0 (X)$ consists of constants.

\item For any $N\geq 0$, $\dim A_N(X)\leq q_{n,N}<\infty$.
All elements of the space $A_N(X)$
are holomorphic Floquet--Bloch functions with quasimomentum $k=0$.

\item There exists a finite family of holomorphic functions
$\{f_j\}$ on $X$ such that all elements of $A_N(X)$ are
polynomials of $\{f_j\}$.
\end{enumerate}
\end{theorem}
\pf Let $k\in \RR^n$ be in $F_{\bar{\partial},\RR}$. This means
that there exists a nonzero analytic function $u(x)$ of Bloch
type, i.e., automorphic with a unitary character $\gamma_k$ of the
group $G$. The absolute value of $u(x)$ is then $G$-periodic and
thus can be pushed down to $M$. Due to the compactness of $M$, it
must attain its maximum. Hence, the function $u(x)$ attains
somewhere on $X$ the maximum of its absolute value. The standard
maximum principle now implies that $u=\mathrm{constant}$. This
means in particular that $k=0$. Hence, (1) holds.

Representation (\ref{E:repr}) of Lemma \ref{L:delta} implies now
that $A_0 (X)$ consists of constants. Indeed, this space is
comprised of analytic functions on $X$ automorphic with respect to
unitary characters of $G$. As we have just shown, every such
automorphic function is automatically constant.

Let us extend the $\bar{\partial}$ operator on $X$ to the elliptic
Dolbeault complex (e.g., \cite[Chapter IV, Examples 2.6 and
5.5]{Wells}):
\begin{equation}\label{E:Dolbeault}
0\rightarrow\mathcal{A}^{0,0}\stackrel{\bar{\partial}}\rightarrow\mathcal{A}^{0,1}
\stackrel{\bar{\partial}}\rightarrow\mathcal{A}^{0,2}\stackrel{\bar{\partial}}\rightarrow\ldots,
\end{equation}
where $\mathcal{A}^{p,q}$ is the sheaf of germs of smooth $(p,q)$-forms on $X$.
The conditions of Theorem
\ref{T:overdet} are now satisfied, which proves the statement (3).

Finally, the implication (3)$\Rightarrow$(4) was established in
\cite{Brud}.\qed

 \mysection{Liouville theorems on combinatorial and
quantum graphs}\label{S:graphs}

Liouville type theorems have also been studied intensively on
discrete objects (graphs, discrete groups) (e.g.,
\cite{Kaim,Marg}). Our technique is insensitive to the local
structure of the base of the covering, and hence it can also be
applied to abelian co-compact coverings of graphs.

Let $\Gamma$ be a countable noncompact graph with a free
co-compact action of an abelian group $G$ of a free rank $n$.
Denote by $\tilde{\Gamma}$ the finite graph $\Gamma /G$. Consider
any periodic finite order difference operator $P$ on $\Gamma$. One
can define all the basic notions: solutions of polynomial growth,
Floquet solutions, quasimomenta, Fermi variety, etc., exactly the
way they were defined before. The same procedure as before leads
to the following result that we state without a proof:
\begin{theorem}\label{T:graph}

\begin{enumerate}
\item If the Liouville theorem of order $N\geq 0$ for the equation $Pu=0$ on $\Gamma$
holds  true, then it holds to any order.

\item In order for the Liouville theorem to hold, it is necessary
and sufficient that the real Fermi surface $F_{P,\RR}$ (i.e., the
set of unitary characters $\chi$ for which the equation $Pu=0$ has
a nonzero $\chi$-automorphic solution) is finite.

\item If the Liouville theorem holds true, then:

\begin{enumerate}
\item Each solution $u \in \mathrm{V}_N(P)$ can be represented as a finite
sum of Floquet solutions of order up to $N$.

\item If $d_N=dim\,\mathrm{V}_N(P)$, then for all $N\in \NN$, we have
$$
d_N\leq d_0 q_{n,N}<\infty\,,
$$
where $q_{n,N}$ is the dimension of the space of all polynomials
of degree at most $N$ in $n$ variables.

\item \label{I:graph item} Assume that for each real quasimomentum $k\in F_{P,\RR}$ the
conditions of Theorem \ref{T:Fl_dimen} are satisfied. Then for
each $N \geq 0$ the dimension $d_N$ of the space $\mathrm{V}_N(P)$
is equal to
$$
   \sum\limits_{k\in F_{P,\RR}}
   r_k\left[  \left(%
\begin{array}{c}
  n+N \\
  N\\
\end{array}%
\right) - \left(%
\begin{array}{c}
  n+N-l_0(k) \\
  N-l_0(k)\\
\end{array}%
\right) \right].
$$
Here $r_k$ and $l_0(k)$ are respectively the multiplicity of the
zero eigenvalue and the order of the first nonzero Taylor term of
the dispersion relation at the point $k \in F_{P,\RR}$. The terms
in this sum are the dimensions of the spaces of
$\lambda_{l_0(k)}$-harmonic polynomials, and polynomially growing
solutions can be described in terms of these polynomials
analogously to Theorem \ref{T:Fl_dimen}.
\end{enumerate}
\end{enumerate}
\end{theorem}

The same can also be done for the so called {\em quantum graphs},
where a differential (rather than difference) equation is considered
on a graph that is treated as a one-dimensional singular variety
rather than a purely combinatorial object (see the details in
\cite{Ku-chapter}-\cite{Ku-graphs2}). A quantum graph $\Gamma$ is a
graph with two additional structures. First of all, the graph must
be {\em metric}, i.e., each edge $e$ must be supplied with a (finite
in our case) length $l_e>0$ and correspondingly with an `arc length'
coordinate. This allows one to do differentiations and integrations
and to define differential operators on $\Gamma$. The simplest ones
are $-\mathrm{d}^2x/\mathrm{d}x^2+V(x)$, where differentiation is
done with respect to the edge coordinates, and $V(x)$ is a
sufficiently nice (e.g., measurable and bounded) potential. In order
to define such an operator, one needs to impose some boundary
conditions at vertices. All such `nice' conditions have been
described (e.g., \cite{Harmer,KoSch,Ku-graphs2}). The simplest
example is the so called Neumann (or Kirchhoff) condition that
requires that the functions are continuous at the vertices and that
the sum of outgoing derivatives at each vertex is equal to zero. A
{\em quantum graph} is a metric graph equipped with a selfadjoint
operator $P$ of the described kind.

Let now $\Gamma$ be a noncompact quantum graph with a free
co-compact isometric action of an abelian group $G$ of a free rank
$n$, and let $\tilde{\Gamma}$ be the compact quantum graph $\Gamma
/G$. We assume that the Hamiltonian $P$ of the quantum graph is
$G$-periodic. The basic notions (solutions of polynomial growth,
Floquet solutions, quasimomenta, Fermi variety, etc.) can again be
defined for this situation. In particular, we consider the
operators $P(k)$, and assume that for each $k\in \mathbb{C}^n$ the
spectrum of $P(k)$ is discrete.
\begin{theorem}\label{T:quantum_graph}
The statements of Theorem \ref{T:graph} hold for the quantum graph
$\Gamma$ under the conditions just described.
\end{theorem}

\mysection{Further remarks} \label{S:remarks}

 \begin{enumerate}
\item
In most parts of the paper we assumed that all the coefficients of
the operators $P$ and $P^{*}$ are $C^\infty$-smooth. This
assumption was made for simplicity only, and in fact one does not
need such a strong restriction (e.g., see the discussion in
\cite[Remark 6.1]{KuPin} and \cite[Section 3.4.D]{Ku}). For
example, for the results of Theorem \ref{T:non-self} to hold, it
is sufficient (but not truly necessary) that the coefficients of
$L$ and $L^{*}$ are H\"{o}lder continuous. It is clear that the
conditions on the coefficients could be significantly relaxed if
the operators were considered in the weak sense, or by means of
their quadratic forms. This does not change the general techniques
of the proofs, which rely on the analytic dependence with respect
to the quasimomenta, rather than on  the particular regularity
properties of the coefficients. One only needs to guarantee the
compactness of the resolvents of the operators $P(\chi)$. More
precisely,  we need that both the operator $P_M$ and its dual
$P_M^{*}$ define Fredholm mappings between the Sobolev space
$H^{m}(M)$ and $L_2(M)$, and this condition can be weakened
further.

\item A stronger statement than Theorem \ref{T:Bloch} holds:
the existence of a subexponentially growing solution also implies
the existence of a Bloch one. This is an analog of the Shnol's
theorem (see \cite{Cycon,Glaz,Shnol}). It is provided in Theorem
4.3.1 of \cite{Ku} for the case of periodic equations in $\RR^n$,
however carrying it over to the case of more general abelian
coverings does not present any difficulty. An analogous theorem
holds for periodic combinatorial and quantum graphs
\cite{Ku-graphs3}.

\item
The interesting feature of the main results of the present paper
is that they relate the Liouville property to the local behavior
of the dispersion relations near the `edges' of the spectrum of
the operator. This threshold behavior is responsible for many
phenomena, in particular homogenization
\cite{Allaire,BS1,BS3,Homog,Conca,Conca2,Jikov}, structure of the
impurity spectrum arising when a periodic medium is locally
perturbed \cite{BirDis1,BirDis2}, Anderson localization, and
others. There are less explored issues of this kind, for instance
the behavior of the Green's function
\cite{Babillot,KuPapPin,Murata,Woess}, integral representations of
solutions of different classes, and finally Liouville theorems.
The last two were also looked upon in our previous paper
\cite{KuPin}. The results of the papers quoted above, as well as
of this paper, deduce properties of solutions from an assumed
spectral edge behavior of the dispersion relation. However, there
are only extremely few results concerning such precise spectral
behavior for specific (or even generic) operators. The notable
exception is the bottom of the spectrum, where much is known
(e.g., \cite{Kirsch,Pr,KobSun} and recent advances in
\cite{BS1,BS2,Shter}). An initial study of the internal edges was
conducted in the interesting paper \cite{KloppRalston}, however
the knowledge in this case remains far from being satisfactory. It
is also appropriate to mention that the study of analytic
properties of the Fermi and Bloch varieties, even for `generic'
operators, happens to be a very hard problem (e.g.,
\cite{GisKnoTru1,GisKnoTru2,KnorTrub}), and there are several
unproven conjectures about the generic behavior of these varieties
(see for example, \cite{AvronSimon,Novikov_VINITI}).

\item As it was mentioned above,
the behavior at the bottom of the spectrum is responsible for
homogenization. Our results on Liouville theorems indicate that in
some sense some kind of ``homogenization'' exists also at the
interior edges of the spectrum. A recent progress has been made
towards precise understanding of the meaning of this
`homogenization' in \cite{Birman-homog} for ODEs and \cite{BS3} for
PDEs.

\item An unexpected and counter-intuitive result of our study of
Liouville theorems is that they depend on the {\bf order} $l_0$ of
the leading term $\lambda_{l_0}$ of the dispersion relation only,
rather than on its full structure. For instance, if $l_0=2$, the
dimensions are the same as for the Laplace operator, even if
$\lambda_{2}$, the Hessian of the dispersion relation, is degenerate
(e.g., the case of the operator $M_t$ with $t=2\sqrt{3}$ in Section
\ref{S:examples}). Normally in homogenization theory, the whole term
$\lambda_{l_0}$ is important.

\item Although our results are complete for the case of non-multiple
spectral edge ($r=1$ in terms of Theorem \ref{T:Fl_dimen} and
\ref{T:Liouville_dim}), in the multiplicity case when $r>1$ we
require that the determinant of the leading term $\lambda_{l_0}$ of
the Taylor expansion is not identically equal to zero. Simple
examples show that without this non-degeneracy condition, the proof
of (\ref{E:dimension}) (and probably the result as well) does not
work anymore. Indeed, suppose that we are dealing with a single
point $k$ of the Fermi surface and that the matrix $\lambda(k)$ is
diagonal, with each diagonal entry $\lambda_{jj}(k)$ having its own
leading term $\lambda_{l_j}(k)$. In this case, it is not hard to
check that the summand in (\ref{E:dimension}) needs to be replaced
by
\begin{equation}\label{E:dimension_corrected}
   \sum\limits_{j}
   \left[
   \left(
   \begin{array}{c}
  n+N \\
  N\\
\end{array}%
\right) - \left(%
\begin{array}{c}
  n+N-l_j \\
  N-l_j\\
\end{array}%
\right) \right],
\end{equation}
in order for the result to hold. One can also see that in the
non-selfadjoint case the non-degeneracy condition in its present
form essentially forces the algebraic and geometric multiplicities
of the zero eigenvalue to coincide. These deficiencies can be
mollified by allowing more general ways of choosing the leading
term $\lambda_{l_0}$, e.g., in the `Douglis--Nirenberg sense'
analogous to the corresponding definition of ellipticity
\cite{Douglis}. However, at the present moment it is not clear to
the authors how to avoid having any non-degeneracy conditions in
the case of a multiple eigenvalue. Fortunately, in most known to
us cases that are interesting for applications, one expects
multiplicity to be absent.

\item The result of Theorem \ref{T:holom} for the case of K\"{a}hler
manifold can be extracted from the elliptic case, by switching the
consideration to harmonic functions. Such result was obtained in
\cite{Brud}. This, however, does not work in general and the
technique presented in this text is needed.

\item It is interesting to mention that the dimension (and in fact the geometry) of the
underlying manifold $X$ does not explicitly enter in our main
results. It definitely influences the geometry of the dispersion
curves and hence the Liouville theorems. However, on the surface
it looks like the group $G=\ZZ^n$ matters more than the manifold.
In particular, one can easily cook up a {\bf two dimensional}
covering $X \mathop{\rightarrow}\limits^{\ZZ^n}M$ with an
arbitrarily large $n$ (take for example the standard
$2$-dimensional jungle gym $JG^2$ in $\mathbb{R}^n$, see
\cite{P-Stroock}). Then in the results that concern the dimensions
of the spaces of polynomially growing solutions, one sees mostly
the influence of $n$. In particular, one can get large dimensions
of these spaces in such a manner. The reason is that, as it has
been mentioned before, our Liouville theorems are of `homogenized'
nature, and being seen from afar, the covering manifold $X$ looks
like $\RR^n$ \cite{Hyperb,Gromov}.
\end{enumerate}

\vskip 3mm
\begin{center}
{\large Acknowledgments} \end{center}
 The
authors express their gratitude to S.~Agmon, M.~Birman,
A.~Brudnyi, L.~Ehrenpreis, P.~Li, V.~Palamodov, V.~Papanicolaou,
R.~Shterenberg, T.~Suslina, M.~Zaidenberg, and N.~Zobin for
information and helpful discussions. Special thanks go to V.~Lin,
discussions with whom have triggered this work.

This research of both authors was supported by Grant No. 1999208
from the United States-Israel Binational Science Foundation (BSF).

The work of P. Kuchment was partially supported by the NSF Grants
DMS 9971674 and 0072248. P. Kuchment expresses his gratitude to
NSF for this support. The content of this paper does not
necessarily reflect the position or the policy of the federal
government of the USA, and no official endorsement should be
inferred. The work of Y.~Pinchover was partially supported by the
Israel Science Foundation founded by the Israeli Academy of
Sciences and Humanities, and by the Fund for the Promotion of
Research at the Technion.


\end{document}